\documentclass[twocolumn,secnumarabic,amssymb, nobibnotes, aps, prapplied]{revtex4-2}

\usepackage[pdftex]{graphicx}
\usepackage{amsmath}
\usepackage{amssymb}
\usepackage{chngcntr}
\usepackage{float}
\usepackage{ifpdf}
\usepackage{subcaption}
\usepackage{caption}
\usepackage{xcolor}
\usepackage{enumitem}
\DeclareMathOperator{\sech}{sech}
\def\eps{\epsilon}
\def\muu{\mu}
\def\sig{\sigma}
\def\sigmam{\sigma_m}
\def\muur{\mu_r}
\def\epsr{\epsilon_r}
\begin{document}

\title{Sum Rule Bounds Beyond Rozanov Criterion in Linear and Time-Invariant Thin Absorbers }

\author{Chen Firestein and Amir Shlivinski}
\email{firestie@post.bgu.ac.il}
\email{amirshli@bgu.ac.il}
\affiliation{School of Electrical and Computer Engineering, Ben-Gurion University of the Negev, Beer Sheva, 84105, Israel}

\author{Yakir Hadad}
\email{hadady@eng.tau.ac.il}
\affiliation{School of Electrical Engineering, Tel-Aviv University, Ramat-Aviv, Tel-Aviv, Israel, 69978}

\begin{abstract}
Dallenbach layer is composed of an absorbing magnetic-dielectric layer attached to a perfect
electric conductor (PEC) sheet. Under linearity and time invariance (LTI) assumptions Rozanov has established analytically a sum-rule trade-off between the absorption efficacy over a predefined bandwidth and
the thickness of the layer, that is the so called Rozanov bound. In recent years several proposals have been introduced to bypass this bound by using non-LTI absorbers. However, in practice, their implementation may be challenging. Here, we expose additional hidden assumptions in Rozanov's derivation, and thus we introduce several new sum rules for LTI layer absorbers that are not covered by the original Rozanov's criterion, and give rise to more relaxed constraints on the absorption limit. We then, demonstrate practical LTI designs of absorbing thin layers that provide absorption beyond the Rozanov's bound. These designs are based on the replacement of the original PEC boundary by various types of penetrable impedance sheet.

\end{abstract}
\maketitle

\section{Introduction}
Wave absorbers play a key role in a variety of electromagnetic \cite{ruck1970radar,Ra'di2015Thin,Landy2008Perfect,qu2022microwave}, acoustic \cite{qu2022microwave,Yang2018integration,Gao2022manipulation,jin2022lightweight,Ryoo2022broadband,Mak2021Going} and optical \cite{Kats2016Optical,Bai2022Boosting,Yao2021Recent} wave systems throughout the entire frequency spectrum.
A canonical type of wave absorber is the Dallenbach layer that is composed of a slab of magneto-dielectric lossy substance backed by a perfect electric conductor (PEC) \cite{ruck1970radar,Dallenbach1938Reflection,Shang2013single_layer,Ye2013Ultrawideband,qu2021conceptual,Yang2017Optimal,Reinert2001potential,Medvedev2022epsilon,rozanov2000ultimate}.
Usually the absorbing bulk is composed of LTI materials, which may be composed of a continuous matter, laminate composite or periodic lattice of conductive, dielectric or magnetic elements at scales that are substantially smaller than the operating wavelength,i.e., metamaterials.
For Dallenbach absorbers, Rozanov has established an analytic bound which relates its absorption efficiency over a predefined bandwidth and the thickness of the layer \cite{rozanov2000ultimate}. A similar relation was later introduced also for the acoustic wave regime at \cite{Yang2017Optimal}.

In recent years, wave engineering using time-varying media has been  utilized as an alternative approach to bypass LTI bounds. Time-varying wave devices allow an additional degree of freedom in comparison to traditional, time invariant wave devices. Therefore, they can potentially overcome some fundamental limitations enforced by the LTI requirements \cite{shlivinski2018beyond,li2019beyond,guo2020improving,solis2020generalization,chen2013broadening,yang2022broadband}.
This approach has been used to design broadband causal reflection-less absorbers \cite{Hayran2021Spectral}, to improve reflection bandwidth for quasimonochromatic signals \cite{li2020temporal}  and in \cite{firestein2021absorption} to bypass Rozanov absorption bound by allowing the characteristics of the layer (permittivity, permeability and electric conductivity) to abruptly or gradually vary with time.
However, the  realization of such a time-varying absorber raises several practical challenges, specifically regarding the ways which the modulation network operates, its complexity, its interference with the wave system, and issues of detection of the impinging wave that should be absorbed.

{The key question that immediately rises is:} {Would it be possible to go beyond Rozanov absorption bound using a passive LTI absorber?} %{\color{red} This key question is discussed in this paper.} %This is the key question of this paper.
{This research question sharpens in light of the well known concept of  perfectly matched layer (PML) that is often used as a means to close the finite computation domain in numerical simulations {and absorb any reflections from the boundaries}. The first implementation of PML, as suggested by Berenger \cite{berenger1994perfectly}, is based on the use of an hypothetical magnetic conductor that is suitably balanced with the electric conductivity. This way, Berenger demonstrates that a perfect absorption can be achieved with no bandwidth limitations using a passive LTI absorber. What is then the fundamental gap between the system considered by Rozanov, which is bounded, and the hypothetical PML layer when backed by a PEC sheet? In the following we strive to bridge this gap, and by doing so we expose the fundamental requirement by any absorber, in order to go beyond Rozanov bound. We emphasize that while the PML layer is hypothetical, the conclusions that we derive in this paper yield to practical realistic designs of passive LTI absorbers that perform beyond Rozanov bound, as discussed in Secs.~\ref{Impedance_Boundary_beyond}} and \ref{Tightness}. Specifically, in these sections we show that by replacing the PEC backing by a properly designed, partially transparent, impedance sheet, the net absorption can be significantly increased compared to what is dictated by Rozanov's bound.

\section{Revisiting Rozanov bound} \label{Rozanov}
\begin{figure}[H]
    \includegraphics[width=8cm]{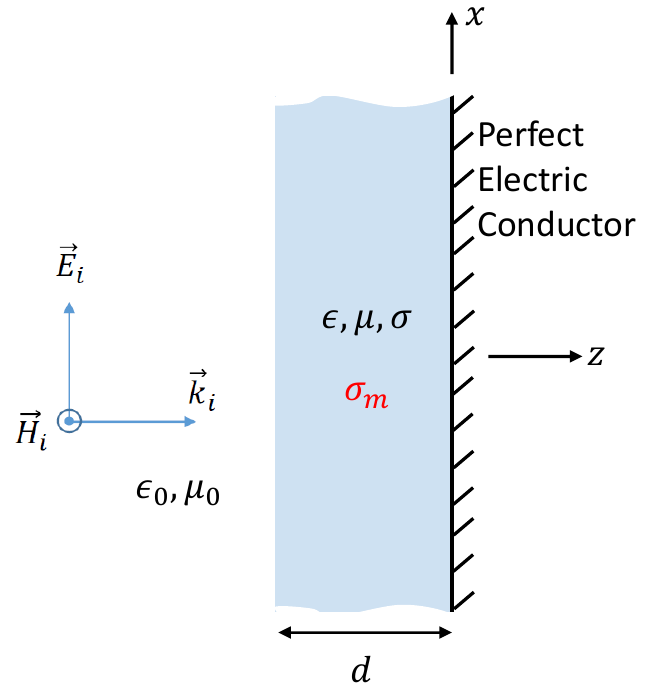}
    \caption{Rozanov's system layout. A normal incidence plane wave impinges a Dallenbach layer that is terminated by a PEC boundary. }
    \label{Roz_PEC_Layout}
\end{figure}

For the sake of self-consistency, in this section we briefly describe the system considered by Rozanov and its bound \cite{rozanov2000ultimate}.
Consider a thin conductive dielectric layer with static (long wavelength) electric parameters $\eps, \muu, \sig$. The layer thickness is $d$ and it is backed by a perfect electric conductor (PEC). See Fig.~\ref{Roz_PEC_Layout} for illustration.
%%%%%%
%%%%%%
%
A plane wave, that is propagating in vacuum ($\eps_0, \muu_0$), impinges the layer normally. For this scenario, Rozanov \cite{rozanov2000ultimate} established a tradeoff between the power absorption efficiency $\tilde{A}(\lambda)=1-|\rho(\lambda)|^2$ given in terms of the reflection coefficient $\rho(\lambda)$, and the bandwidth
\begin{equation}
\label{Rozi_Origi}
{\frac{1}{2}\left| \int_{0}^{\infty} \!\!\ln \left[1-\tilde{A}(\lambda)\right]\! \,d\lambda \right|=}\left|\int_0^{\infty}\ln|\rho(\lambda)| d\lambda \right| \le 2\pi^2\muur d.
\end{equation}
Eq.~\eqref{Rozi_Origi} implies that the tradeoff between the absorption efficiency and the bandwidth depends only on the static (long wavelength) relative permeability $\muur=\muu/\muu_0$ and the layer thickness $d$. This result originates from the long wavelength behaviour of the reflection coefficient in this case
\begin{equation}\label{Eq_Rozanov_Ref}
\rho(\lambda) \sim - 1 + j\frac{4\pi\muur d}{\lambda} \mbox{ as } \lambda\rightarrow\infty.
\end{equation}
For detailed derivation of this sum rule and additional sum rules please refer to Appendices \ref{AppendixA},~\ref{AppendixB}.

\section{What can we learn from the PML concept in order to go beyond Rozanov's bound?} \label{magnetic_conductive_beyond}
In this section we repeat the derivation of Rozanov bound for a system that includes an additional hypothetical static magnetic conductance ($\sigmam$), as assumed by Berenger \cite{berenger1994perfectly}.
%By doing so, we introduce an additional degree of freedom to the setup. We show that the resulting reflection coefficient approximation in the long wavelength range is substantially different than the one shown in Eq.~\eqref{Eq_Rozanov_Ref}, and consequently, as opposed to the setup considered by Rozanov \cite{rozanov2000ultimate}, it is impossible to derive a direct bound on $|\rho(\lambda)|$. {\color{red} However}, absorption beyond Rozanov bound is achievable, and even ideal absorption for any bandwidth. Thus addressing the apparent contradiction with the PML concept.
By doing so, we introduce an additional degree of freedom to the setup. We show that the resulting reflection coefficient approximation in the long wavelength range is substantially different than the one shown in Eq.~\eqref{Eq_Rozanov_Ref}.  Consequently, Rozanov's bound, Eq.~\eqref{Rozi_Origi}, is invalid. Moreover, it is impossible to derive a direct bound on $\ln |\rho(\lambda)|$.  Instead, absorption beyond Rozanov bound is achievable, and even ideal absorption for any bandwidth. Thus we conclude, that controlling the large wavelength behaviour of the reflection coefficient is the key to go beyond Rozanov's bound. The detailed derivation is given below.

%We consider a Dallenbach layer with magnetic conductive substance terminated by a PEC boundary (see Fig.~\ref{Fig_Sketch} (a)) and derive analytic bounds on the reflection coefficient. We provide numerical and experimental demonstrations that support our main claim of this section - wave absorption beyond Rozanov's bound can be achieved by considering magnetic conductive substances which are mathematically modeled in a TL to series resistance (as opposed to electric conductive sheets that are modeled as shunt admittance).

%\subsection{Analytic relation}
The reflection coefficient at the interface between the vacuum and the absorbing medium reads,
\begin{equation}\label{reflection}
\rho(\lambda)=\frac{Z_{\rm{in}}(\lambda)-\eta_0}{Z_{\rm{in}}(\lambda)+\eta_0},
\end{equation}
where $Z_{\rm{in}}(\lambda)=Z_0(\lambda)\tanh(\gamma(\lambda)d)$ is the input impedance at the interface between the Dallenbach layer and the surroundings, $Z_0(\lambda)$ is the characteristic impedance and $\gamma(\lambda)$ is the complex propagation term (see Appendix \ref{AppendixA}) and $\eta_0$ is the free space wave impedance. Interestingly, for long wavelength, $\left(|\lambda|\rightarrow \infty \right)$, it asymptotically behaves as
\begin{subequations}
\label{mod_rho}
\begin{equation}\label{rho_sigma_sigmam}
\rho(\lambda)\sim-1 + A+j2\pi B/\lambda
\end{equation}
where
\begin{equation}\label{Eq_A}
A=\frac{2\kappa}{1+\kappa}\mbox{, with }\kappa=\frac{1}{\eta_0} \sqrt{\frac{\sigmam}{\sig}}\tanh\left({\sqrt{\sig\sigmam}}d\right)
\end{equation}
and
\begin{equation}\label{Eq_B}
     B=c_0\frac{d\left(\frac{\muu}{\sigmam}+\frac{\eps}{\sig}\right) \left(\frac{\sigmam}{\eta_0}-\eta_0 \sig \kappa^2\right)+\left(\frac{\muu}{\sigmam}-\frac{\eps}{\sig}\right)\kappa}{(1+\kappa)^2}
\end{equation}
\end{subequations}
where $\eps=\eps_0\epsr$ ($\epsr$ is the static relative permittivity), {similarly $\sigma$ and $\sigma_m$ are the static electric and magnetic conductivities, respectively,} and $c_0=1/\sqrt{\eps_0 \muu_0}$ is the free space wave speed.
The constant $A$ is recognized as the transmission coefficient $\left(T\right)$ into the absorbing layer (which serves as a load) at long wavelength, i.e., $A=T(|\lambda|\rightarrow \infty)$.
It can be observed  (see Appendix \ref{AppendixA}) that by setting the magnetic conductivity $\sigmam$ to zero, Eq.~\eqref{rho_sigma_sigmam} reduces to Rozanov's long wavelength approximation for the reflection coefficient \cite{rozanov2000ultimate} that is in Eq.~\eqref{Eq_Rozanov_Ref}.

%While at first glance the approximations in Eq.~\eqref{Eq_Rozanov_Ref} and Eq.~\eqref{rho_sigma_sigmam} seem similar, the presence of the extra term $A$ in Eq.~\eqref{rho_sigma_sigmam} implies a striking consequence. This is because under these circumstances, $|\int_{0}^{\infty} \ln |\rho(\lambda)| \,d\lambda |$ by itself \emph{cannot be bounded from above} and therefore the maximal absorption is not limited.
While at first glance the approximations in Eq.~\eqref{Eq_Rozanov_Ref} and Eq.~\eqref{rho_sigma_sigmam} seem similar, the presence of the extra term $A$ in Eq.~\eqref{rho_sigma_sigmam} implies that  $|\int_{0}^{\infty} \ln |\rho(\lambda)| \,d\lambda |$ by itself \emph{cannot be bounded from above} and therefore the maximal absorption is not limited.  Instead,  we may readily derive a bound on a modified expression (see Appendix \ref{AppendixB}), giving
\begin{equation}\label{Generalized_Relation}
    \int_{0}^{\infty} \ln |\rho(\lambda)-A| \,d\lambda \geq - \pi^2B.
\end{equation}
{This bound implies} that the maximal absorption is intimately connected with the coefficient $A$.
Obviously, it can be observed in Eq.~\eqref{Generalized_Relation} that \emph{the static parameters themselves set the bound over the entire real wavelength axis} without any additional frequency dependent contributions.
Also, by setting $\sigmam\rightarrow0$ in Eq.~\eqref{Generalized_Relation}  Rozanov's bound is recovered.

The presence of static magnetic conductance, $\sigmam$, provides an additional  degree of freedom that enables going beyond the limitation suggested by Rozanov regarding the  maximal absorption efficacy (that can be viewed as a tradeoff between absorption and signal's bandwidth).
Specifically, the inclusion of $\sigmam$ implies a change of the long wavelength reflection coefficient as assumed in Eq.~\eqref{rho_sigma_sigmam}. First, in the zero order term when $A\neq0$, and second, in the first order term as shown by comparing Eq.~\eqref{mod_rho} %Eq.~\eqref{rho_sigma_sigmam}
and Eq.~\eqref{Eq_Rozanov_Ref}.

Ideally \emph{infinite} magnetic conductance, i.e., the so called, artificial  magnetic conductor (AMC) can be synthesized in a narrow frequency band as an high impedance surface. This in fact was one of the first practical applications of metamaterials \cite{Sievenpiper1999High-impedance, Feresidis2005Artificial, Zhang2003Planar, Engheta2002Thin, Tretyakov2003Thin, Simms2005Thin, Kamgaing2003Anovel}.  However, their physical structure renders them to exhibit long wavelength characteristics as of a PEC backed structure and therefore they comply with Rozanov's bound.
As opposed to that, unfortunately, the use of \emph{static} \emph{finite} magnetic conductance is generally impossible for airborne waves using natural materials nor metamaterials  (For guided waves it can be synthesized e.g., a transmission line with lossy conductors can be treated as effectively having magnetic conductance. See the discussion in the supplementary material file \cite{SM} and in Appendix~\ref{AppendixE}).
Nevertheless, although static finite magnetic conductance is hypothetical, the results in this section have practical significance since they shed light on what should be done in order to go beyond Rozanov bound with a passive LTI system. Specifically, we note that in order to breach the bound, we need to find ways to alter the asymptotic, long wavelength, behaviour of the reflection coefficient $\rho(\lambda)$ compare to Eq.~\eqref{rho_sigma_sigmam}. As we show below, this can actually be done, by replacing the PEC boundary with a penetrable impedance sheet.
Careful designs may allow a substantial improvement of the absorption efficiency and bandwidth.

%%%%%%%%%%%%%%%%
\section{Generalization of Rozanov's bound - the impedance boundary condition}\label{Impedance_Boundary_beyond}
%%%%%%%%%%%%%%%%
%\textcolor {red} {As opposed to electric conductive substances, natural magnetic conductive materials do not exist in nature and there is no realization of an effective magnetic conductive bulk with a controllable, finite magnetic conductivity.} This technological barrier implies that the description in Sec.~\ref{magnetic_conductive_beyond} can be readily used for TL absorbers rather than physical 3D wave absorbers. Nevertheless, it provides insights regarding other LTI venues to go beyond the Rozanov absorption bound.
%Here we describe an alternative approach where the Dallenbach absorber is terminated by an arbitrary non-opaque impedance (see Fig.~\ref{Fig_Sketch} (b)) with no magnetic conductivity.
%%We tackle Rozanov's assumption on the reflection coefficient at large wavelength and show that although both reflected and transmitted waves exist (due to the penetrability of the boundary) supreme absorption results can be obtained, since the reflection coefficient is differ than the one used by Roznaov.
%In such a system there are both reflected and transmitted waves due to the penetrability of the boundary. However, careful design allows an increased absorption performances in comparison to Rozanov's bound due to invalidity of the assumption on the reflection coefficient at large wavelength (see Eq.~\eqref{Eq. Rozanov Ref}).

\begin{figure}[H]
         \centering
    \begin{subfigure}[b]{\columnwidth}
         \centering
         \includegraphics[width=8cm]{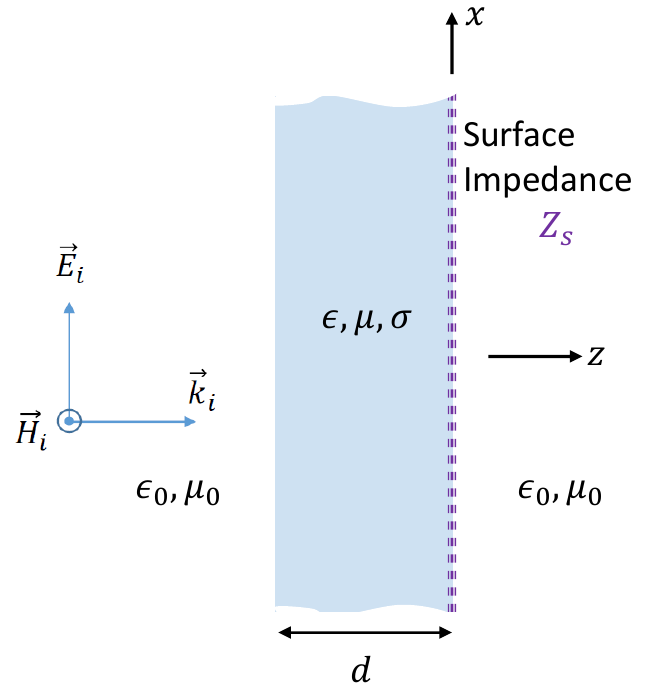}
         \caption{}
         \label{Dallenbach_Impedance_FS}
    \end{subfigure}
    %\hfill
    \begin{subfigure}[b]{\columnwidth}
         \centering
         \includegraphics[width=8cm]{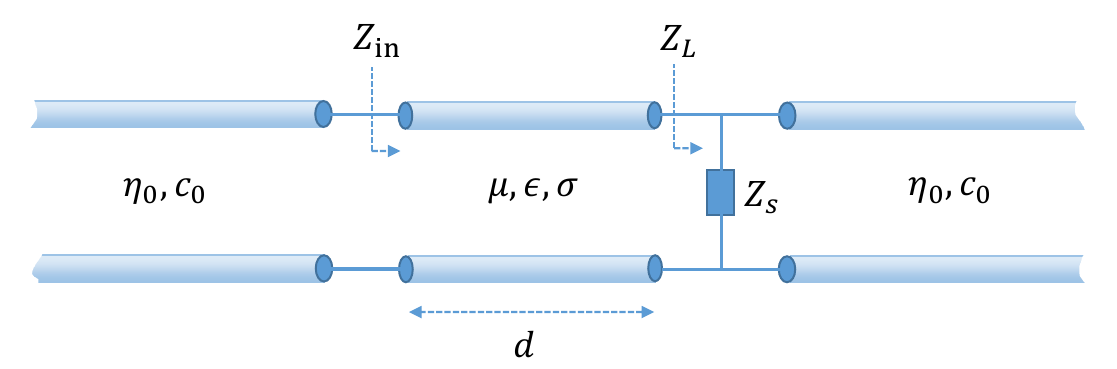}
         \caption{}
         \label{Dallenbach_Impedance_TL}
    \end{subfigure}
    \caption{Description of the problem. (a) An incident wave propagates (left to right) in a semi-infinite medium towards a lossy Dallenbach layer with thickness $d$ that is loaded by a non opaque impedance sheet followed by another semi-infinite medium. (b) TL model of the problem.}
    \label{Dallenbach_Impedance_a_b}
\end{figure}

Here we describe an alternative approach to bypass Rozanov's bound by replacing the absorber's terminating sheet by a non-opaque impedance sheet \emph{with no magnetic conductivity} (see in Fig.~\ref{Dallenbach_Impedance_a_b}).
%Here we describe an alternative approach {\color{red} to bypass Rozanov's bound} where the Dallenbach absorber is terminated by an arbitrary non-opaque impedance sheet \emph{with no magnetic conductivity} {\color{red} instead of the PEC sheet} {\color{red} (see the physical realzition in Fig.~\ref{Dallenbach_Impedance_FS} and the corresponding transmission line model in Fig.~\ref{Dallenbach_Impedance_TL})}.
%
%
%
%\subsection{Large wavelength approximation of the reflection and transmission coefficients}
The sheet's impedance has the following generalized form with the long wavelength approximation, %($\lambda\rightarrow \infty$)}
%%
%\begin{equation}\label{Impedance_Sheet}
%    Z_{\rm{s}}(\lambda)=\sum_{n=-\infty}^{\infty} A_n \left(2j\pi c_0/\lambda\right)^n \sim A_k \left(2j\pi c_0/\lambda\right)^k
%\end{equation}
%%
%\underrightarrow{(\eta-\epsilon)}
%\begin{equation}\label{Impedance_Sheet}
%    Z_{\rm{s}}(\lambda)=\sum_{n=k}^{\infty} A_n \left(2j\pi c_0/\lambda\right)^n \sim A_k \left(2j\pi c_0/\lambda\right)^k
%\end{equation}
\begin{equation}\label{Impedance_Sheet}
    Z_{\rm{s}}(\lambda)=\sum_{n=k}^{\infty} A_n \left(2j\pi c_0/\lambda\right)^n \xrightarrow{\lambda \to \infty} A_k \left(2j\pi c_0/\lambda\right)^k
\end{equation}
where the $A_k$'s are real positive constants. Note that $k=0$ indicates a resistive loading while $k \gtrless 0$ indicates a reactive loading.
%
%where $k$ denotes the lowest power with non-zero coefficient, i.e. $A_n=0$ for $n<k$.
The corresponding load impedance, $Z_L$ , at large wavelength as seen when looking into the interface where the impedance sheet is located {(see in Fig.~\ref{Dallenbach_Impedance_TL})} reads,
\begin{equation}\label{Load_imp}
    Z_{L}(\lambda)=Z_{\rm{s}} \parallel \eta_0 \sim \left(\frac{1}{A_k \left(2j\pi c_0/\lambda\right)^k}+\frac{1}{\eta_0}\right)^{-1},
\end{equation}
whereas the input impedance as observed at the interface between the absorbing material and its surroundings reads,
\begin{equation}\label{Input_imp_1}
\begin{array}{ccc}
  Z_{\rm{in}}(\lambda)=Z_0(\lambda)\frac{Z_L(\lambda)+Z_0(\lambda)\tanh(\gamma d)}{Z_0(\lambda)+Z_L(\lambda)\tanh(\gamma d)}.
 % \sim \\ \\ A_k \left(2j\pi c/\lambda\right)^k +2j\pi c \mu d/\lambda.
\end{array}
\end{equation}
%
%These impedances are readily expressed via a transmission line model as shown in Fig.~\ref{Dallenbach_Impedance_TL}.

%%%%%%
%%%%%%
%
The long wavelength behaviour of $Z_{\rm{in}}$ depends on the asymptotic long wavelength behaviour of $Z_L$ in Eq.~\eqref{Impedance_Sheet} which is discussed in the following for the two cases $k \geq 1$ and $k \leq 0$.
%Following Eq.~\eqref{Load_imp}, the long wavelength behaviour of $Z_L$ depends on whether $k \geq 1$ or  $Z_{\rm{s}}$ has a positive ($k\geq1$) or non-positive ($k\leq0$) reactance:
%
\begin{itemize}[leftmargin=*]
%\emph{positive reactance.}
\item \emph{positive reactance.} This case implies $k\geq1$, thus $Z_L \sim A_k \left(2j\pi c_0/\lambda\right)^k$ and consequently, following Eq.~\eqref{Input_imp_1}
%The input impedance as observed at the interface between the absorbing material and its surroundings reads,
%
\begin{equation}\label{Input_imp}
\begin{array}{ccc}
  Z_{\rm{in}}(\lambda) \sim A_k \left(2j\pi c_0/\lambda\right)^k +2j\pi c_0 \muu d/\lambda.
\end{array}
\end{equation}
It can further be observed that for $k>1$ the first order approximation of the input impedance is $Z_{\rm{in}} \sim 2j\pi c_0 \muu d/\lambda$. Using Eq.~\eqref{reflection} (see also Appendices \ref{AppendixA},\ref{AppendixB}) the reflection and transmission coefficients revert back to the form that was used by Rozanov, i.e. $\rho\sim-1+4\pi j \muur d/\lambda$ and $T \sim O(1/\lambda^2)$.
The physical implication of this result is that for $k>1$ it is impossible to design a Dallenbach absorber that improves the absorption performances with respect to Rozanov bound.
%of a Dallenbach absorber that is terminated by a PEC (for which Rozanov's bound was derived).
%with improved absorption performances in comparison to Rozanov's bound, i.e. Dallenbach absorber that is terminated by a PEC boundary.

For the case of $k=1$, the surface impedance $Z_{\rm{s}}$ is that of a pure inductor, therefore, setting $A_1=L$ (H) in Eq.~\eqref{Load_imp}. %In this work, the surface inductance represents the total inductance of the loading circuit within a single unit cell which is equivalent to the total inductance of the impedance sheet due to its square dimensions.
Repeating the derivation in Appendix \ref{AppendixA} but for this case,  it follows that the long wavelength approximation of the reflection and transmission  coefficients are given by,
\begin{equation}\label{R_T_k_1}
\begin{split}
\rho(\lambda) &\sim -1+4j\pi\muur\left[ d +\frac{L}{\left(\muu_0\muur\right)}\right]\lambda^{-1}
\\
T(\lambda) &\sim \frac{4j\pi  L}{\muu_0 \lambda}.
\end{split}
\end{equation}
%
%\begin{equation}\label{R_T_k_1}
%\begin{array}{ccc}
%  \rho(\lambda) &\sim -1+4j\pi\mu_r\left[ d +\frac{L}{\left(\mu_0\mu_r\right)}\right]\lambda^{-1}
% \\
% T(\lambda) &\sim \frac{4j\pi  L}{\mu_0 \lambda}.
%\end{array}
%\end{equation}
%
%Note that by setting $L=0$ in Eq.~\eqref{R_T_k_1}, the inductive boundary degenerates into a PEC, therefore the reflection coefficient takes the original form  that was used by Rozanov in Eq.~\eqref{Eq. Rozanov Ref}.
The modified expression of the reflection coefficient in Eq.~\eqref{R_T_k_1}, differs by a single linear term in ${L}/{\left(\muu_0\muur\right)}$ from that of Rozavov's PEC terminated Dallenbach absorber which by comparison to Eq.~\eqref{Eq_Rozanov_Ref} further leads to a modified Rozanov-type bound, as we show next.
Conservation of power in this passive LTI absorbing system dictates that
\begin{equation}\label{Power_relation}
    |\rho\left(\lambda\right)|^2+|T(\lambda)|^2+\tilde{A}(\lambda)=1,
\end{equation}
where $\tilde{A}(\lambda)$ is the absorption efficacy.
%
%where $\tilde{A}(\lambda)$ is the normalized energy absorption efficiency. CHEN: should be power, not energy.
Based on Eq.~\eqref{Power_relation}, a generalized absorption bound is derived by
\begin{equation}
\label{Inductive_bound}
\begin{aligned}
    &\frac{1}{2}\left| \int_{0}^{\infty} \!\!\ln \left[1-\tilde{A}(\lambda)\right]\! \,d\lambda \right|\\
    \\&=\frac{1}{2}\left| \int_{0}^{\infty}\!\!\ln \left[|\rho(\lambda)|^2+|T(\lambda)|^2\right]\! \,d\lambda\right|  \\ \\
    &\leq
    {\left|\int_{0}^{\infty} \!\!\ln |\rho(\lambda)|\! \,d\lambda \right|}
%    \frac{1}{2}\left|\int_{0}^{\infty} \!\!\ln |\rho(\lambda)|^2\! \,d\lambda \right|
    \leq 2\pi^2\muur\left[d+\frac{L}{\muu_0\muur}\right],
\end{aligned}
\end{equation}
where the first equality in Eq.~\eqref{Inductive_bound} follows from Eq.~\eqref{Power_relation}, the next inequality follows from the fact that $0 \le |\rho(\lambda)|,|T(\lambda)| < 1$ with $\left[|\rho(\lambda)|^2\!+\!|T(\lambda)|^2\right]<1$ and the last inequality is derived similarly to Eq.~\eqref{Rozi_Origi} (see Appendix ~\ref{AppendixB}). Note that the corresponding Rozanov's type bound for a PEC terminated absorber ($L=0$), reads $\frac{1}{2}\left| \int_{0}^{\infty} \!\ln \left[1-\tilde{A}(\lambda)\right]\! \,d\lambda \right| \leq 2\pi^2\muur d$ (see Eq.~(\ref{Rozi_Origi})).

The discussion so far demonstrated the counterintuitive result that the use of a Dallenbach absorber with penetrable termination (instead of Rozanov's PEC bound) may provide a venue for obtaining absorber with improved absorption but with a transmission beyond the absorber. This result will be explored and demonstrated in Sec.~\ref{Tightness} with specific numerical details.

\item \emph{non-positive reactance case.} In this case $k \le 0$ which implies for a resistive/capacitive impedance sheet with the long wavelength approximation of the load impedance,
\begin{equation}\label{Load_imp_cap_res}
  Z_{L}(\lambda) \sim \begin{cases}
  (\frac{1}{A_0}+\frac{1}{\eta_0})^{-1}  & k=0, \\
  \eta_0-j2\pi c_0  \eta_0^2 /(A_{-1}\lambda)  & k=-1, \\
  \eta_0 & k<-1.
\end{cases}
\end{equation}

At large wavelengths the transmitted wave {beyond the absorber} is no longer negligible, i.e. $|\rho|\not\gg|T|$. Therefore, the large wavelength absorption is affected by both the reflection and transmission coefficients which make it impossible to bound the absorption as in Eqs.~\eqref{Rozi_Origi} and~\eqref{Inductive_bound}. However, as we prove next, the absorption at extremely large wavelengths is limited.

{For the capacitive type impedance sheet,}
at extremely large wavelengths the impedance of {the loading}
%capacitor loading
is substantially larger in comparison to the impedance of the free space, therefore $Z_L=\eta_0$ (see in, Eq.~\eqref{Load_imp}). Here, we use a $0^{\rm{th}}$ order approximation of $\rho(\lambda), T(\lambda)$ to bound from above the absorption coefficient ($\tilde{A}$) under this limit.
Using Eq.~\eqref{Input_imp_1}, the input impedance as observed at the input of the absorber is given by $Z_{\rm{in}}\sim \eta_0/\left(1+\eta_0\sig d\right)$ and the reflection ($\rho$) and transmission ($T$) coefficients read,
\begin{align}\label{RT_approx_0}
    \rho \sim \frac{-\eta_0 \sig d}{2+\eta_0 \sig d},\hspace{0.5cm} T \sim \frac{2}{2+\eta_0 \sig d}.
\end{align}
Applying $\rho$ and $T$ with Eq.~\eqref{Power_relation} gives
\begin{equation}\label{function_x}
    1-\tilde{A}=|\rho|^2 +|T|^2=\frac{(\eta_0 \sig d)^2+4}{(\eta_0 \sig d+2)^2}
\end{equation}
which has a global minimum at $\sig_{\rm{opt}}=2/(\eta_0 d)$  with maximal large wavelength absorption of $\tilde{A}(\lambda \rightarrow \infty)=0.5$.
%Denoting $x=\eta_0 \sig d$ where $x\geq 0$ and applying Eq.~\eqref{Power_relation} with Eq.~\eqref{RT_approx_0} leads to,
%\begin{align}\label{function_x}
%    1-\tilde{A}(x)=|\rho(x)|^2 +|T(x)|^2=\frac{x^2+4}{(x+2)^2}.
%\end{align}
%The function $1-\tilde{A}(x)$ has a global minimum at $x=2$, therefore the maximal large wavelength absorption is $\tilde{A}(\lambda \rightarrow \infty)=0.5$ at $\sig_{\rm{opt}}=2/(\eta_0 d)$.

In Sec.\ref{Resistive_Capacitive_Boundary_Absorption} we demonstrate numerically that this approach makes it possible to go beyond Rozanov's absorption performances.

\end {itemize}
%\subsection{Inductive boundary condition - absorption bounds and parametric study}\label{Inductive_Boundary_Absorption}

\section{The tightness of the bound: inductive case}\label{Tightness}
The bound in Eq.~\eqref{Inductive_bound} is general for any passive LTI inductively loaded penetrable Dallenbach type absorber. This bound reflects the somewhat counterintuitive fact that the net absorption may be increased, compare with Rozanov's setup that is backed by PEC, by using a partially transparent layer that enables transmission trough it.
%
%Furthermore, it establishes that previous observation that even in the presence of a transmitted wave $T(\lambda)$, beyond the absorber, and with a careful design, an improved absorption, as compared to the standard Dallenbach case, can be obtained.
%
Furthermore, from the expression in Eq.~\eqref{Inductive_bound} it should also be noted, as for Rozanov's bound \cite{rozanov2000ultimate} and other sum rules \cite{Bode1945Network,Gustafsson2020Upper,Schab2020Trade,Abdelrahman2022How}, that \emph{only the static behaviour of the impedance sheet determines the overall absorption bound.}
Since Eq.~\eqref{Inductive_bound} suggests that the bound we found for the total absorption in the presence of an inductive impedance sheet is larger (less stringent) than Rozanov's bound. The immediate question is, is it possible to find practical designs that are tight to the new bound? Below we provide an affirmative answer to this question and demonstrate that indeed the new bound is tight.

In order to address this question, we define the tightness of the bound using  the following optimization problem
\begin{equation}
\begin{split}
\label{Tight_par}
&\tau(L)\triangleq \max_{\eps,\sig}\frac{\frac{1}{2}\left| \int_{0}^{\infty} \!\ln \left[1-\tilde{A}(\lambda)\right]\! \,d\lambda \right|}{2\pi^2\muur\left[d+L/\left(\muu_0\muur\right)\right]}
 \\[2ex]
&\textrm{subject to: Kramers-Kronig}  \\
\end{split}
\end{equation}
where $\tau(L)\in \left[0,1\right]$ for $L\geq0$.
In the denominator, we utilize the expression for the bound taken from Eq.~\eqref{Inductive_bound}. On the other hand, in the numerator, we present the achieved absorption for a design that has undergone optimization. The optimization process involves varying the values of permittivity and conductivity ($\epsilon, \sigma$), while keeping the inductance $L$ fixed at predetermined values.
Note that the infinite integral in Eq.~(\ref{Tight_par}) cannot be performed numerically. Instead, in any numerical calculation we resort to integration over a finite wavelength range [$\lambda_1,\lambda_2$]. Therefore, we define below an auxiliary tightness parameter
\begin{equation}
\begin{split}
\label{Tight_par_aux}
&\tau_{[\lambda_1,\lambda_2]}(L)\triangleq \max_{\eps,\sig}\frac{\frac{1}{2}\left| \int_{\lambda_1}^{\lambda_2} \!\ln \left[1-\tilde{A}(\lambda)\right]\! \,d\lambda \right|}{2\pi^2\muur\left[d+L/\left(\muu_0\muur\right)\right]}
 \\[2ex]
&\textrm{subject to: Kramers-Kronig}  \\
\end{split}
\end{equation}
Note that the \emph{integrands} in Eqs.~(\ref{Tight_par},\ref{Tight_par_aux}) are positive for any value of $\lambda$ and for any physical choice of layer parameters, satisfying Kramers-Krong relations, i.e., layer that is passive, causal, and linear time-invariant. As a result,
\begin{equation}\label{tau_ineq}
\tau_{[\lambda_1,\lambda_2]}\le \tau\le 1.
\end{equation}
The second inequality originates from Eq.~(\ref{Inductive_bound}).
Consequently, while in principle, in order to demonstrate the tightness of the bound we need to solve the optimization problem in Eq.~(\ref{Tight_par}) and show that $\tau$ approaches 1 from below. In light of Eq.~(\ref{tau_ineq}), it is in fact enough to solve the optimization problem in Eq.~(\ref{Tight_par_aux}) and show that $\tau_{[\lambda_1,\lambda_2]}$ approaches 1 from below.
Note that the second optimization problem, is the only one that can be performed computationally. In this calculation we assume that the layer is nonmagnetic i.e., with $\mu_r$ set to 1, and with thickness $d=0.4 \left(m\right)$. In addition, we assume that the parameters $\epsilon$ and $\sigma$ are practically constant in the waveband $[\lambda_1,\lambda_2]=[0.01,3\times10^7]$, which is a reasonable assumption for most dielectric materials since it implies the material resonance should occur above, say, $300$(GHz) which is about 10 times higher then the highest frequency in the integration interval, namely,  $f=3\times10^8(\rm{m/s})/0.01(\rm{m})=30$GHz.
Taking into account the above considerations we assure that our calculation approach is  subject to Kramers-Kronig relations (See \cite{Jackson1999Classical}, and additional discussion in appendix \ref{AppendixD}).

%
%
%Addressing the absorption bound tightness is carried out numerically, and by doing so the concept of carefully (optimized) design will be demonstrated. To that end, consider a non-magnetic Dalllenbach layer ($\mu_r=1$) with thickness $d=0.4 \left(m\right)$. Recalling that in the absorber's design there are additional two parameters that are not \emph{implicitly} involved in the expression for the bound, those are the permittivity $\epsilon$ and the electric conductivity $\sigma$ (see in Eq.~\eqref{Inductive_bound}). For a given $L$, define the tightness ($\tau$) of the bound by the optimization over the range of the possible $\epsilon$ and $\sigma$ parameters
%%
%\begin{equation}
%\begin{split}
%\label{Tight_par}
%&\tau(L)\triangleq \max_{\eps,\sig}\frac{\frac{1}{2}\left| \int_{0}^{\infty} \!\ln \left[1-\tilde{A}(\lambda)\right]\! \,d\lambda \right|}{2\pi^2\muur\left[d+L/\left(\muu_0\muur\right)\right]}
% \\[2ex]
%&\textrm{subject to:} \quad \{\eps>0, \sig>0\} \\
%\end{split}
%\end{equation}
%where $\tau(L)\in \left[0,1\right]$ for $L\geq0$. Following the optimization in Eq.~\eqref{Tight_par},

\begin{figure}[H]
\centering
\includegraphics[width=8cm]{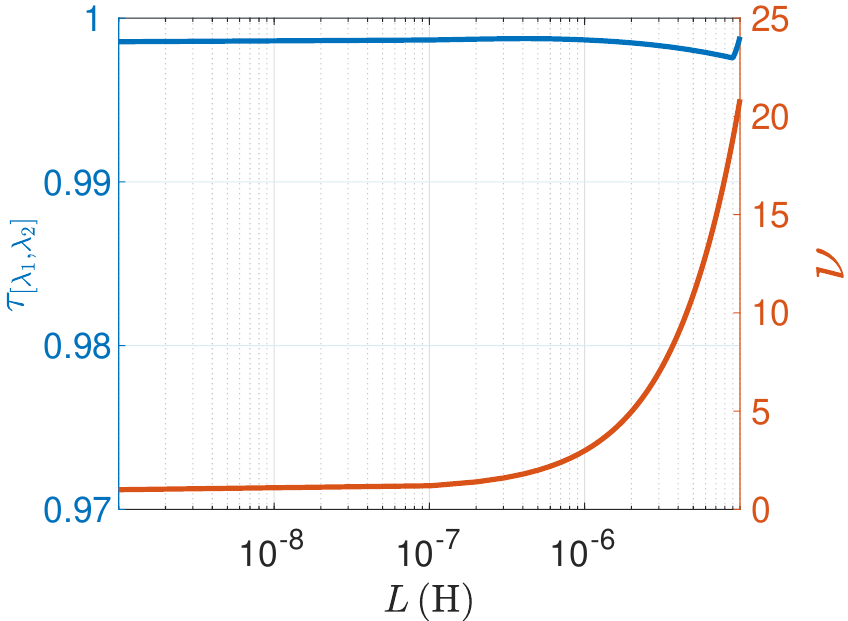}
\caption{The ratio of the integration result of the optimal design and the inductive bound, $\tau$, of Eq.~\eqref{Tight_par} in blue line and the corresponding value with Rosanov's bound, $\nu$ of Eq.~\eqref{Ratio1} in red line.}
    \label{Tighten_Bound}
\end{figure}

The optimization results are shown by the blue line in Fig.~\ref{Tighten_Bound} that depicts $\tau_{[\lambda_1,\lambda_2]}$ as a function of the inductance $L$. It can readily be observed  that the bound in Eq.~\eqref{Inductive_bound} is tight, with $0.99<\tau_{[\lambda_1,\lambda_2]} \le\tau\le 1$, for reasonable values of inductance ($L \leq 10 \left(\mu \rm{H}\right)$).
Furthermore, to highlight the comparative advantage in relation to the maximum achievable performance according to Rozanov's bound, we present a comparison between the new bound and Rosanov's bound. This is given by the expression below,
\begin{equation}\label{Ratio1}
    \nu = \frac{\frac{1}{2}\left| \int_{0}^{\infty} \!\ln \left[1-\tilde{A}(\lambda)\right]\! \,d\lambda \right| }{2\pi^2\muur d},
\end{equation}
%%%
%%%%%%
and the corresponding results are shown by the red line in Fig.~\ref{Tighten_Bound}.
%%%%%%
From the figure we observe that for moderate inductance values, $L>0.1 \left(\mu \rm{H}\right)$, the new absorption bound becomes significantly  better than Rosanov's bound. As opposed to that, for smaller inductance  values, $L<0.1 \left(\mu \rm{H}\right)$, where the inductive sheet becomes effectively a short circuit, i.e., a PEC boundary, the new absorption bound reduced to Rozanov's bound. Note that in this case,  the total scattered field is composed of solely the reflected field, and the transmitted field is negligible in this case. Thus, quite surprisingly, with a proper design, we show that  by allowing transmittance through the absorbing layer, the net absorbance can be largely increased.

\begin{figure}[H]
\centering
    \includegraphics[width=8cm]{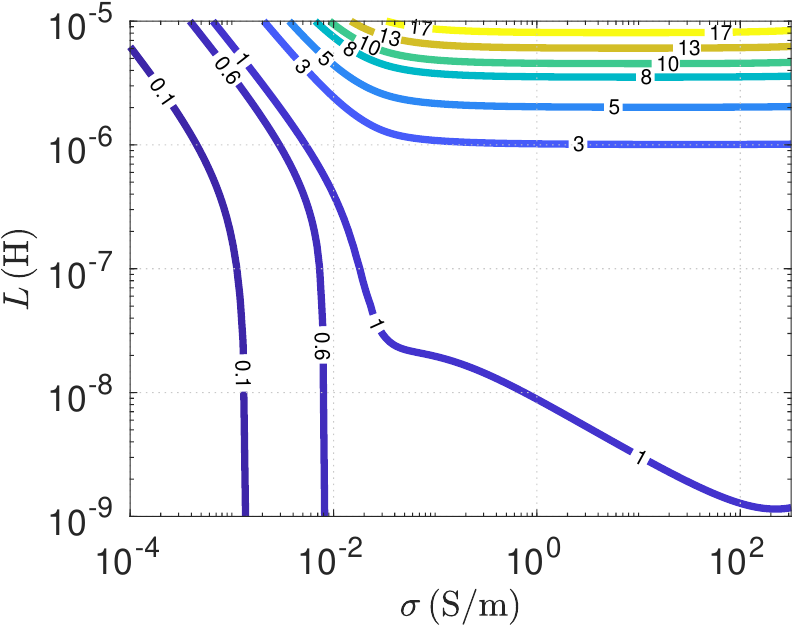}
    \caption{A contour plot (``isolines'') of $\nu$, the ratio of the absorption integration of the inductive impedance boundary and the Rozanov bound.}
    \label{Contour_L_sigma}
\end{figure}

Next, the interplay between $L$ and $\sigma$ is explored for a given $\epsilon_r$ and $\mu_r$, Fig.~\ref{Contour_L_sigma} depicts a contour plot (``isolines'') of $\nu$ as a function of the sheet inductance, $L$, and $\sig$ for $\muur=1, \epsr=1.5$.
%%%%%%
%%%%%%
It can be observed in Fig.~\ref{Contour_L_sigma} that there is a wide range of parameters that indeed give improvement ($\nu>1$). This can be explained by noting that at long wavelength a thin absorbing layer provides additional series inductance of $\mu_0\mu_rd$.
%is inductive by its nature, and this inductance $\mu_rd$ determines the bound by Rozanov.
Therefore, by introducing an additional inductive surface impedance, the effective layer inductance increases and therefore the bound relaxes as seen in Eq.~(\ref{Inductive_bound}) compare to Eq.~(\ref{Rozi_Origi}). However, albeit being intuitive, this explanation is not complete since an inductive boundary does not necessary guarantee practical enhancement in the absorption, as can be viewed by the domains where $\nu<1$, therefore perform poorly in comparison to an optimal Rozanov's absorber. The reason for such cases should be directly associated with the fact that the surface is partially transparent. This observation makes our result even less intuitive since it hints on a gentle balance between the absorption and transmittance processes that is dictated by the additional degree of freedom, in the form of the included inductive impedance surface.

\subsection{Practical design - inductive boundary}\label{Inductive_Boundary_Example}
\begin{figure}[H]
     \centering
     \begin{subfigure}[b]{0.49\columnwidth}
         \centering
         \includegraphics[width=\textwidth]{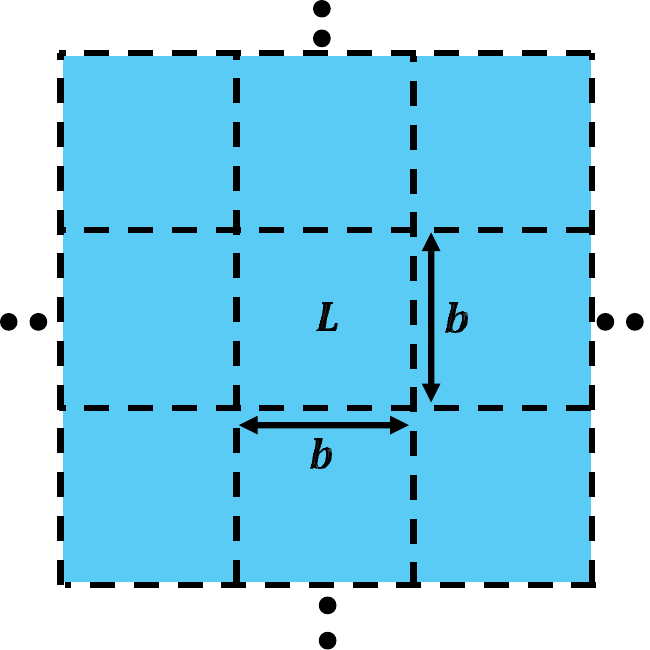}
         \caption{}
         \label{Coax_unit_1}
     \end{subfigure}
     \hfill
     \begin{subfigure}[b]{0.49\columnwidth}
         \centering
         \includegraphics[width=\textwidth]{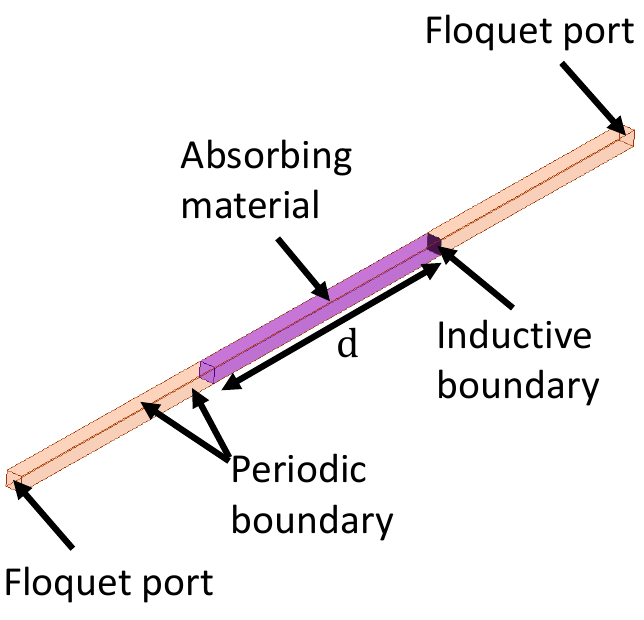}
         \caption{}
         \label{Reflection_coefficient_1}
     \end{subfigure}
\\  \vspace{10pt}
     \centering
     \begin{subfigure}[b]{0.99\columnwidth}
         \centering
         \includegraphics[width=\textwidth]{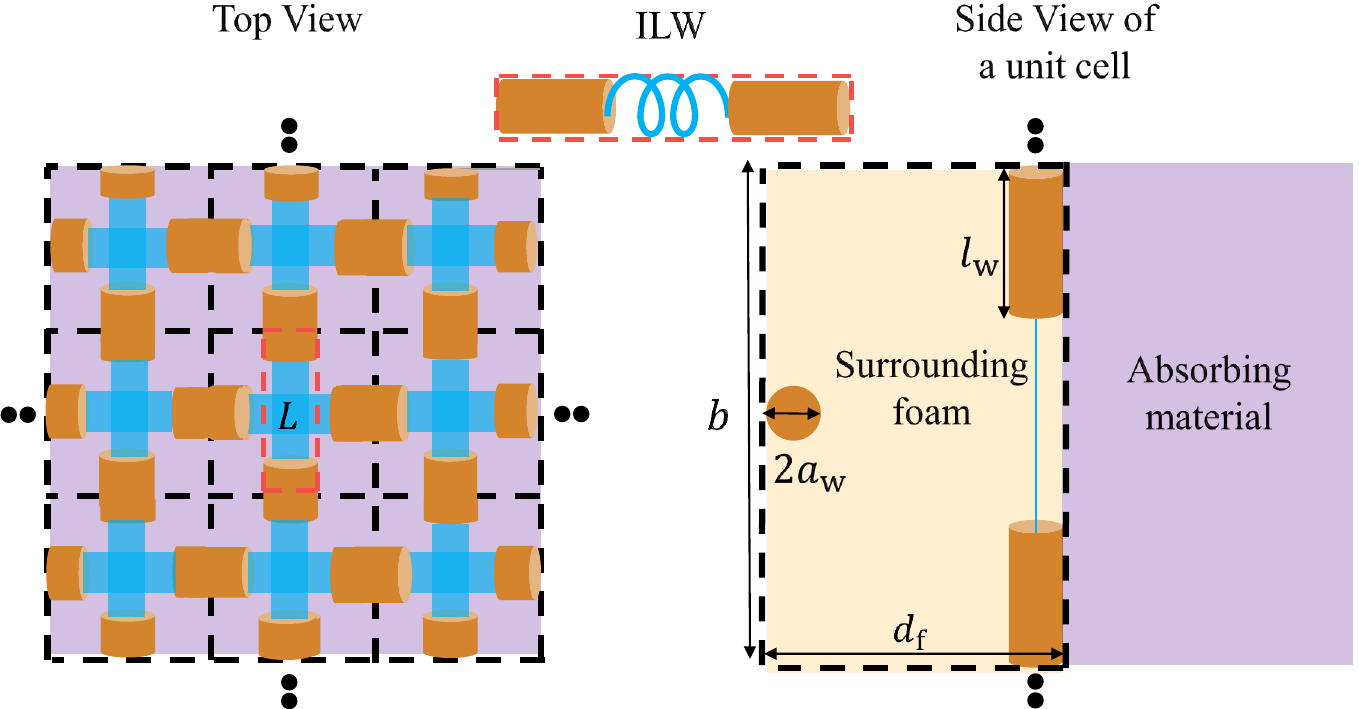}
         \caption{}
         \label{Coax_unit_2}
     \end{subfigure}
\\  \vspace{10pt}
     \hfill
     \begin{subfigure}[b]{0.49\columnwidth}
         \centering
         \includegraphics[width=\textwidth]{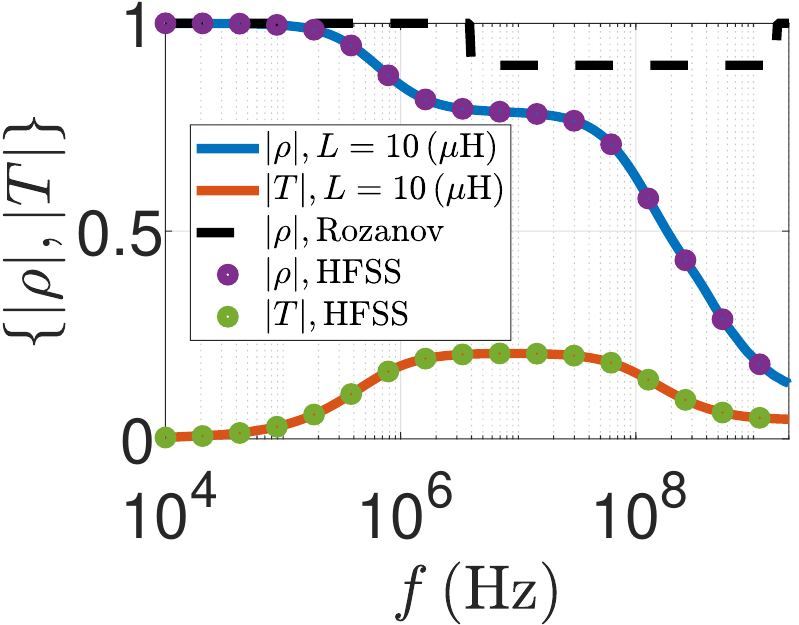}
         \caption{}
         \label{Reflection_coefficient_2}
     \end{subfigure}
     \hfill
     \centering
     \begin{subfigure}[b]{0.49\columnwidth}
         \centering
         \includegraphics[width=\textwidth]{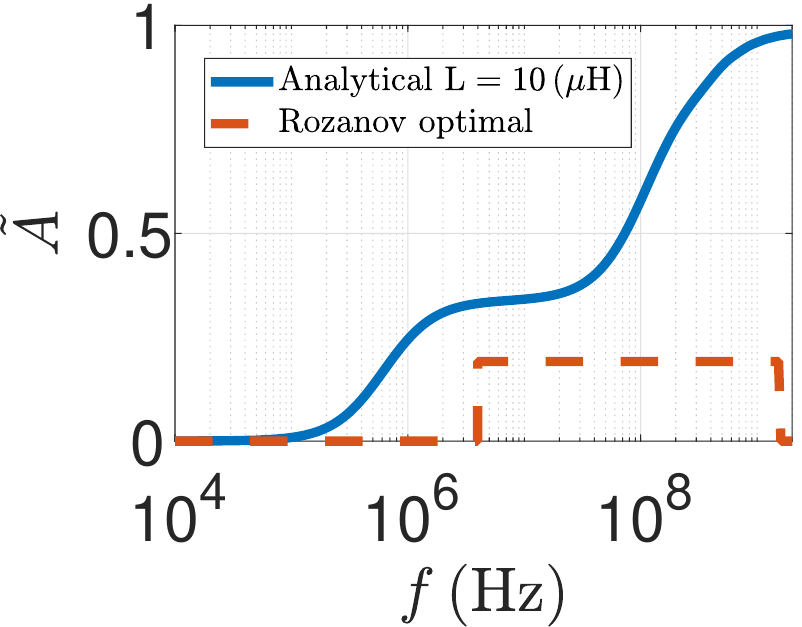}
         \caption{}
         \label{Coax_unit_3}
     \end{subfigure}

        \caption{ Design of wave absorber with an inductive impedance boundary. (a) The boundary is implemented with inductors (blue) that occupy the entire unit cell cross section (theoretical impedance sheet).
        (b) HFSS simulation setup. The periodic structure is excited by a Floquet port where only the fundamental mode is propagating and being reflected, transmitted or absorbed.
        (c) Square grid of ILWs serves as an inductive impedance sheet that is capable of operating under both polarizations (due to the existence of horizontal and vertical loaded wires).
        (d) Reflection (analytical - blue, simulated - purple circles) and transmission (analytical - red, simulated - green circles) coefficients as a function of the operating frequency. In addition, optimal results (Rozanov bound) of a PEC backed absorber are presented in dashed black line. (e) Absorption coefficient as a function of frequency (blue) compared to an optimal design of a PEC backed absorber (red dashed line).}
        \label{Practical_Design}
\end{figure}

A practical design of an ultrawideband wave absorber with inductive boundary condition is presented $(k=1)$. The surroundings is assumed with $\eps_0=8.85\times 10^{-12} \left(\rm{F}/m\right),\muu_0=1.2566\times10^{-6}\left(\rm{H}/m\right)$. The absorber thickness is $d=0.4 \left(\rm{m}\right)$ with parameters $\epsr=1.5,\muur=1,\sig=0.05\left(\rm{S}/m\right)$ and terminated by a non-opaque inductive impedance sheet with inductance of $L=10 \left(\mu \rm{H}\right)$.
The inductive impedance has been realized in two different periodic ways (unit cell dimension is taken as $b=20\left(\rm{mm}\right)$), {the first in} Fig.~\ref{Practical_Design}(a) depicts an ideal impedance sheet, where the inductor occupies the entire cross section of the unit cell.  Figure~\ref{Practical_Design}(b) illustrates the simulation layout that was implemented in HFSS \cite{HFSS}. {The second realization in} Fig.~\ref{Practical_Design}(c) depicts a practical implementation of the impedance sheet by PEC inductively loaded wires (ILW). The length of each wire is $2l_{\rm{w}}=10 \left(\rm{mm}\right)$ distributed equally between two arms with thickness of $2a_{\rm{w}}=2 \left(\rm{mm}\right)$ which leads to a tiny parasitic inductance that is negligible in comparison with the loading inductor \cite{Grover2004Inductance}. The design includes both vertical and horizontal ILWs which enable the absorber to operate at both polarizations of the incident wave. The wires are separated by a distance of $d_{\rm{f}}=4 \left(\rm{mm}\right)$ surrounded by a standard foam material \cite{Foam_Mat}. The figure captures both top and side views.
Figure~\ref{Practical_Design}(d) depicts the analytically calculated reflection (blue) and transmission (red) as a function of the operating frequency, while the corresponding simulated HFSS results are presented in purple and green circles, respectively with an excellent agreement (the simulations were performed for both the theoretical impedance and the wired sheet, yielding similar results with a tiny relative error, less than $2\%$ on average, between the two polarizations). In addition, the optimal Rozanov's behavior of the reflection coefficient, i.e. {$|\int_{0}^{\infty} \ln |\rho(\lambda)| \,d\lambda | \le 2\pi^2 \muur d$}, with uniform reflection in the frequency band $f \in \{4, 1600\} \left(\rm{MHz}\right)$ is depicted in dashed black line. Figure~\ref{Practical_Design}(e) described the absorption coefficient as a function of frequency, where it can be observed that the performances of the impedance inductive sheet absorber are enhanced in comparison to the optimal PEC case, i.e. Rozanov's bound.
The improvement can be seen at any frequency of operation, i.e. from extreme low frequencies up to higher frequencies. We stress that by increasing the minimal frequency (reducing the maximal wavelength) of operation in Rozanov's bound, the absorption performances may be improved at higher frequencies.
%HFSS simulation was performed for both the theoretical impedance and the wired sheet, yielding similar results with a tiny relative error between the two polarizations ($<2\%$ on average throughout the frequency axis).

%%%%%%%%%%%%%%%%%%%%%%%%%%%%
%%%%%%%%%%%%%%%%%%%%%%%%%%%%

\subsection{Other types of impedance sheets}\label{Resistive_Capacitive_Boundary_Absorption}
\begin{figure}[H]
     \centering
     \begin{subfigure}[b]{0.49\columnwidth}
         \centering
         \includegraphics[width=\textwidth]{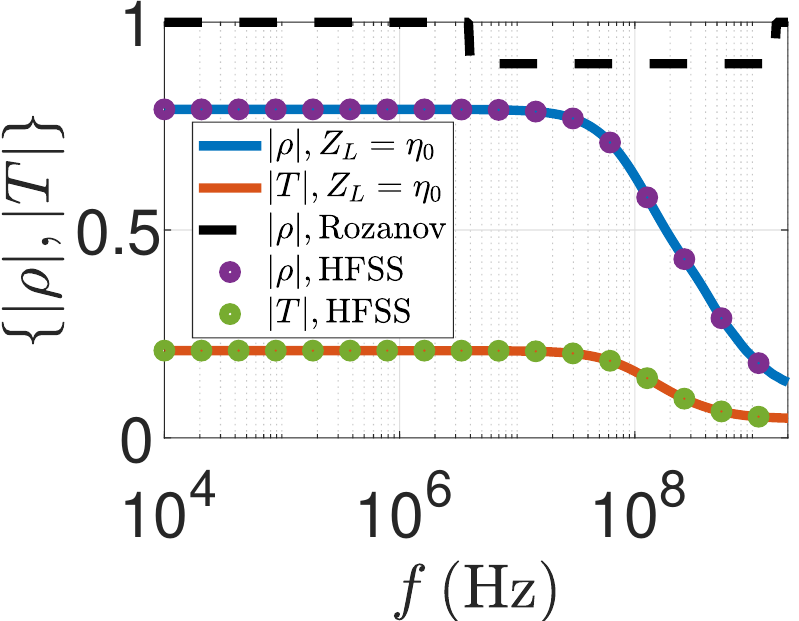}
         \caption{}
         \label{Coax_unit_4}
     \end{subfigure}
     \hfill
     \begin{subfigure}[b]{0.49\columnwidth}
         \centering
         \includegraphics[width=\textwidth]{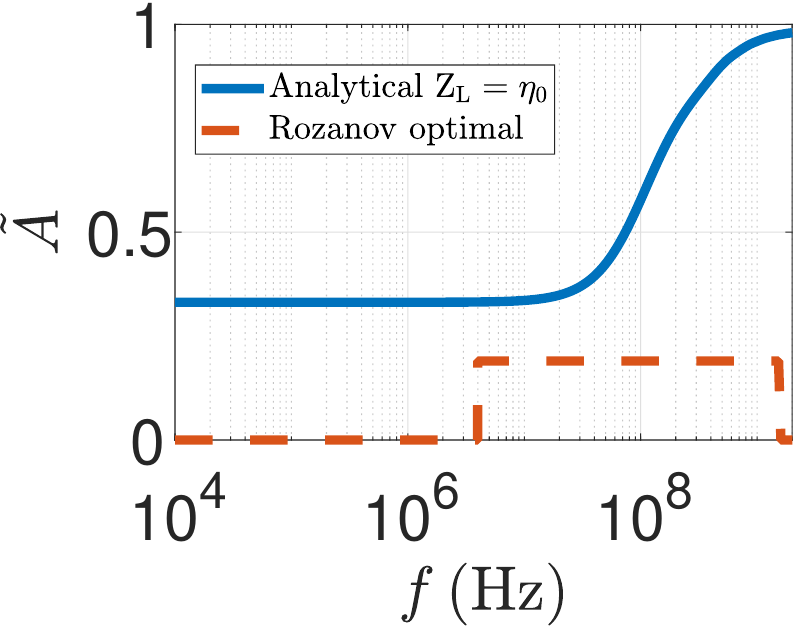}
         \caption{}
         \label{Reflection_coefficient_4}
     \end{subfigure}
     \\
       \begin{subfigure}[b]{0.49\columnwidth}
         \centering
         \includegraphics[width=\textwidth]{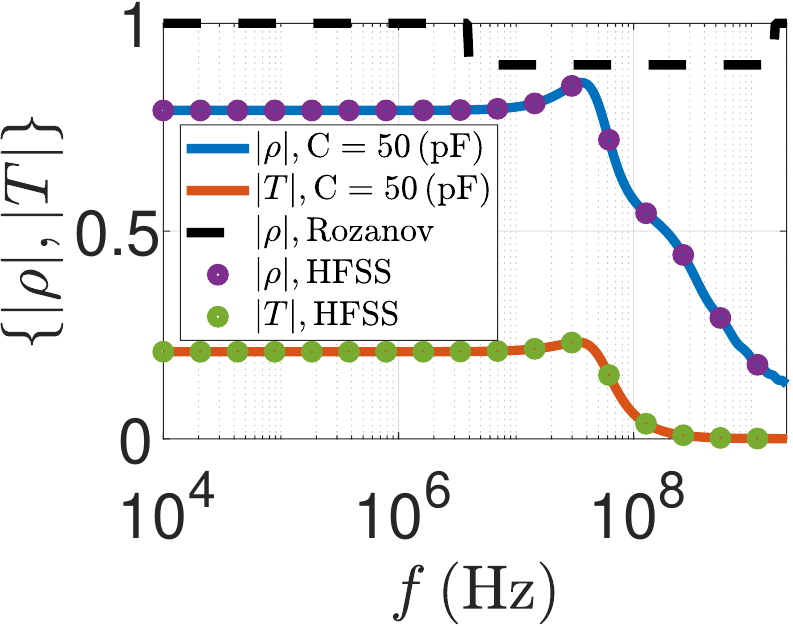}
         \caption{}
         \label{Coax_unit_5}
     \end{subfigure}
     \hfill
     \begin{subfigure}[b]{0.49\columnwidth}
         \centering
         \includegraphics[width=\textwidth]{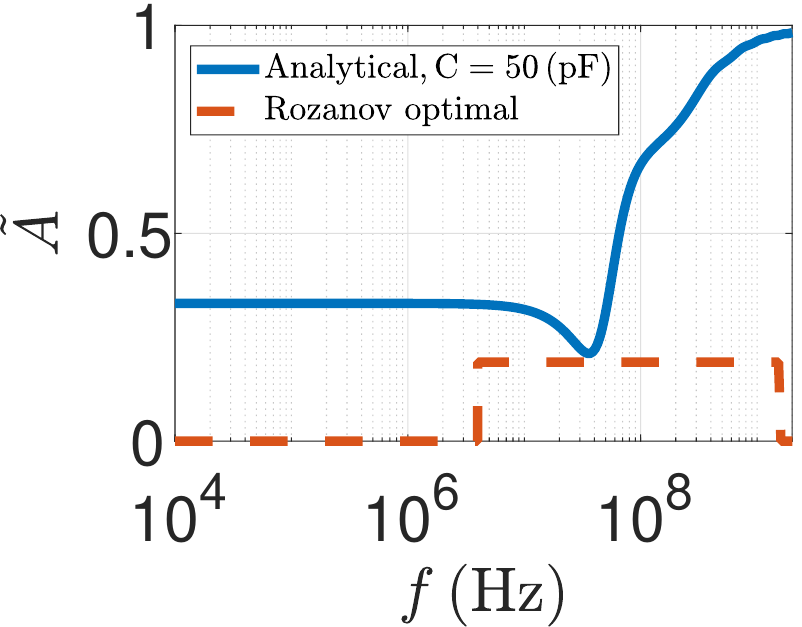}
         \caption{}
         \label{Reflection_coefficient_5}
     \end{subfigure}
        \caption{Reflection, transmission and absorption results of a resistive boundary, $Z_L=\eta_0$ (an open absorber) are presented in (a,b) and described for a capacitive boundary with $C= 50 (\rm{pF})$ in (c,d). (a,c) Reflection (analytical - blue, HFSS simulation - purple circles) and transmission (analytical - red, HFSS simulation - green circles) coefficients as a function of the operating frequency. Optimal results according to Roznaov's bound are presented in dashed black line. (b,d) Absorption coefficient as a function of frequency for the resistive, capacitive impedances (blue),respectively. An optimal design of a PEC backed absorber is presented in a red dashed line.}
        \label{Practical_Design_Resistor_Capacitor}
\end{figure}

Here, we consider the case of a Dallenbach absorber terminated at one end by a non-opaque resistive [capacitive] impedance sheet ($k\leq 0$). As previously discussed, such an absorber allows the existence of both reflected and transmitted waves. In fact, the resistive [capacitive] case has a unique feature in comparison to the previous cases, when considering extremely large wavelength the transmission coefficient is no longer negligible ($|\rho|\not\gg|T|$) thus cannot be neglected when deriving {an absorption sum rule} as performed in Eq.~\eqref{Inductive_bound}. {Therefore no analytic relation, similar to Eq.~\eqref{Inductive_bound} of the inductive case can be found.} %Therefore, there is no analytic relation that we obtain similarly to Eqs.~\eqref{Generalized_Relation}, \eqref{Inductive_bound}.
However, the absorption performance can be analytically explored and numerically  verified by electromagnetic simulations (HFSS). Figure ~\ref{Practical_Design_Resistor_Capacitor}(a) depicts the reflection and transmission coefficients as a function of the operating frequency (analytical and HFSS simulation results) for the resistive case ($Z_L=\eta_0$), where Fig~\ref{Practical_Design_Resistor_Capacitor}(b) depicts the absorption coefficient ($\tilde{A}$) in comparison with these of the optimal Rozanov's bound with uniform reflection in the frequency band $f \in \{4, 1600\} \left(\rm{MHz}\right)$.
Figure ~\ref{Practical_Design_Resistor_Capacitor}(c,d) describe the corresponding results  for the capacitive case with $C= 50(\rm{pF})$, $Z_{\rm{s}}= 1/(j2\pi f C)$.

\begin{figure}[H]
     \centering
     \begin{subfigure}[b]{0.475\columnwidth}
         \centering
         \includegraphics[width=\textwidth]{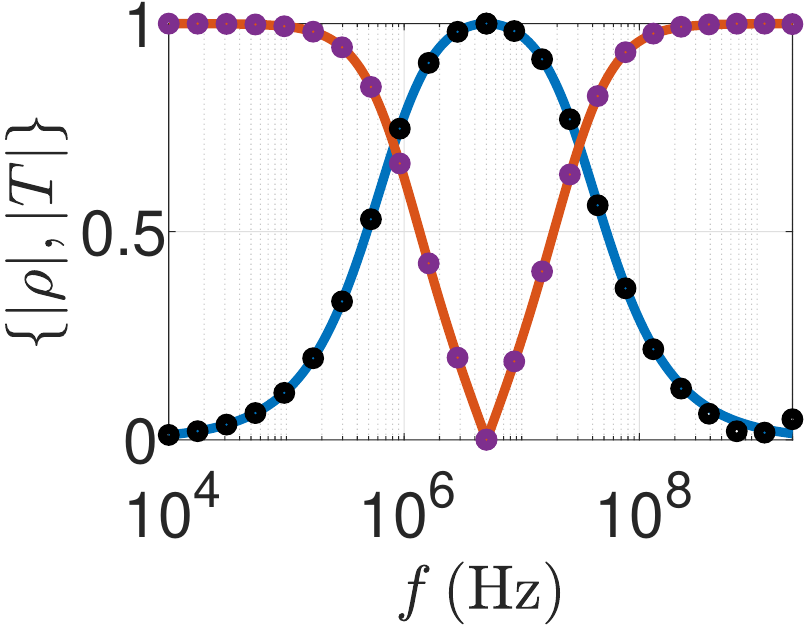}
         \caption{}
         \label{Coax_unit_4}
     \end{subfigure}
     \hfill
     \begin{subfigure}[b]{0.505\columnwidth}
         \centering
         \includegraphics[width=\textwidth]{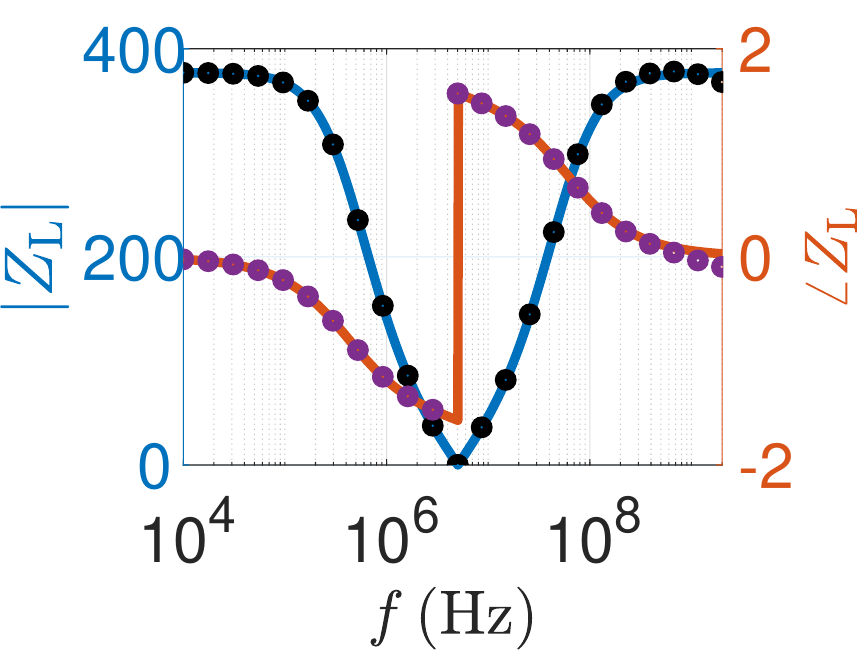}
         \caption{}
         \label{Reflection_coefficient_4}
     \end{subfigure}
     \\
       \begin{subfigure}[b]{0.49\columnwidth}
         \centering
         \includegraphics[width=\textwidth]{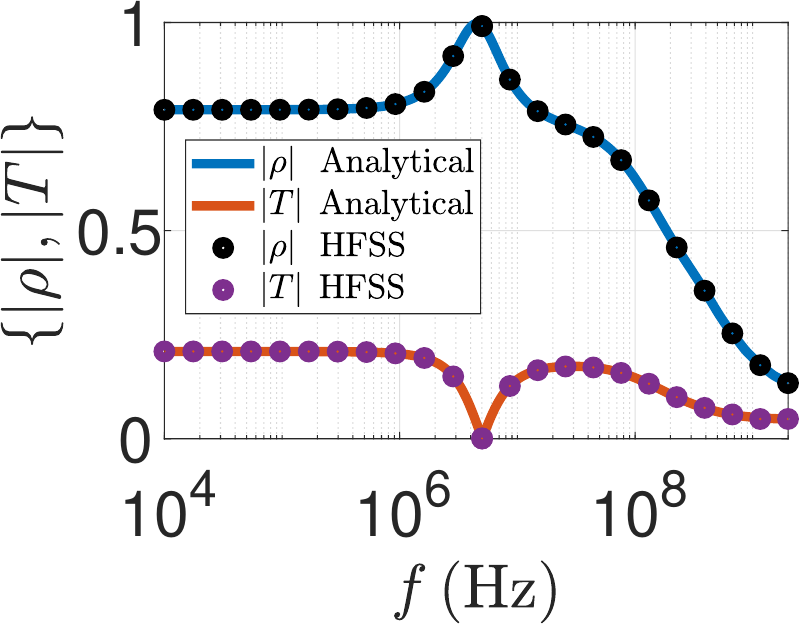}
         \caption{}
         \label{Coax_unit_5}
     \end{subfigure}
     \hfill
    \begin{subfigure}[b]{0.49\columnwidth}
         \centering
         \includegraphics[width=\textwidth]{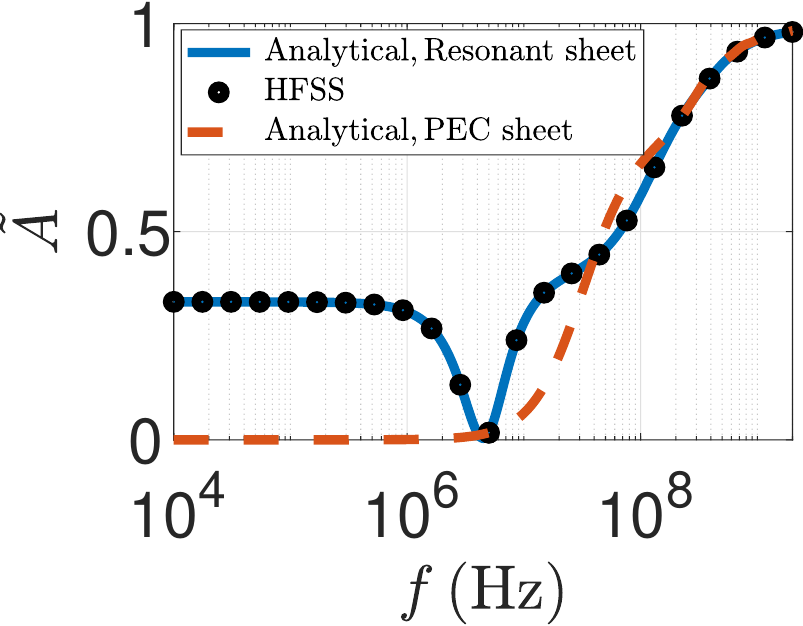}
         \caption{}
         \label{Coax_unit_5}
     \end{subfigure}
     \caption{Resonator design. (a) Reflection (blue - analytical, black dots - simulated) and transmission (red - analytical, purple dots - simulated) coefficients of a resonant impedance sheet (capacitive-inductive sheet) in free space without the absorbing substance. The layout of the impedance sheet is identical to the one described in Sec.~\ref{Inductive_Boundary_Example} where the inductive loading has been replaced by a series connection of a capacitor $C=10^{-9} \left(F\right)$ and an inductor $L= 10^{-6} \left(H\right)$. (b) Loading impedance $Z_L$ (complex value).(c) Reflection and transmission coefficients of the absorbing structure terminated by the capacitive-inductive impedance sheet. (d) Absorption coefficient of a Dallenbach absorber terminated by the resonant sheet (blue - analytical , black dots - simulated) and by a PEC backed sheet (dashed red).  }
        \label{Resonantor}
\end{figure}

Following the anlytic derivation in Sec.~\ref{Impedance_Boundary_beyond} and the numerical results in Fig.~\ref{Practical_Design_Resistor_Capacitor} it can be noted that $\frac{1}{2}\left| \int_{0}^{\infty} \!\ln \left[1-\tilde{A}(\lambda)\right]\! \,d\lambda \right| \not < \infty$.
This non-trivial result can be understood by observing a discrete transmission-line model (that is composed of a periodic arrangement of unit cells containing series inductive-type impedances and shunt capacitive-type admittance, see \cite[Fig.~(S1(b))]{SM}.
%This non-trivial result can be readily explained by observing the TL equivalence (see Sec.~\ref{magnetic_conductive_beyond}).
At the low frequency limit, the shunt admittance (which is analogous to electric conductivity) is the dominant component within each unit cell, since the inductors impedance vanishes while the impedance of the shunt capacitors is extremely large. Therefore, the lossy TL physically behaves as a lumped resistor that is connected in parallel to a load impedance representing the free space ($\eta_0$) implying a non-zero scattering (reflection, transmission) and absorption. It can be observed in Fig.~\ref{Practical_Design_Resistor_Capacitor} that at low frequencies the capacitive loading impedance behaves similarly to the resistive loading, since the impedance of the capacitor is dominant with respect to the free space impedance, $\left|Z_{\rm{s}}\right| \gg \eta_0$.
On the other hand, at large frequencies the opposite relation holds ($\left|Z_{\rm{s}}\right| \ll \eta_0$), thus the boundary behaves similarly to PEC. It can be observed in Figs.~\ref{Practical_Design}-\ref{Practical_Design_Resistor_Capacitor} that at large frequencies (small wavelengths), the absorption is negligibly affected by the loading impedance, since the thickness of the Dallenbach layer is substantially larger with respect to the operating wavelength. At the extreme case of $f \rightarrow \infty$, i.e $d/\lambda \rightarrow \infty$, the absorbing layer is infinitely long, thus rendering the bounding sheet redundant. This physical observation further underlines the discussion above that improved low frequency absorption is possible by altering the boundary condition (similar observation were made in \cite{Mak2021Going} for an acoustical realization).
%%%%%%%%%%%%%%%%%%%%%%%%%%%%
The numerical examples presented above (and also in Sec.~\ref{Inductive_Boundary_Example}) state that using an inductive, capacitive or resistive termination instead of a standard, typical, PEC sheet results in an improved absorption performance due to reduction of the total scattering at low frequencies (large wavelengths).
%However, the behaviour of a PEC backed absorber (that was considered by Rozanov) at certain frequencies may be significant for some applications such as for XXX. Therefore, based on the layout described in Sec.~\ref{Inductive_Boundary_Example}, we design a resonant impedance sheet that operates as a PEC at the resonant frequency.
However, there is also a transmitted field beyond the absorbing layer that is mandatorily blocked for any Dallenbach absorber which obeys the Rozanov's bound. This blocking, which occurs  across the entire frequency range, due to the PEC, may be too restrictive for some applications, such for example if blocking is needed only in a defined frequency range. In such cases, one can use an impedance loading that exhibits shorts circuit behaviour only in this frequency. To that end, a series type resonant impedance loading that is tuned to that frequency range can be used. This is demonstrated in Fig.~\ref{Resonantor}.
Here, the wires are loaded by a capacitor ($C=1 \left(\rm{nF}\right)$) and an inductor ($L= 1 \left(\rm{\mu H}\right)$) connected in series, turned to resonate at $f_{\rm{res}}\sim 5.03 \left (\rm{MHz}\right)$.
As an initial step, we present in Figs.~\ref{Resonantor} (a)-(b) the behaviour of the resonant sheet itself, i.e. in a free space without the absorbing Dallenbach layer.
Figure~\ref{Resonantor}(a) depicts the reflection (blue - analytical, black dots - simulated) and transmission (red - analytical, purple dots - simulated) coefficients as a function of the operating frequency, while Fig.~\ref{Resonantor} (b) depicts the corresponding loading impedance, $Z_L$. Figures~\ref{Resonantor} (c)-(d) describe the reflection, transmission and absorption coefficients of a Dallenbach layer terminated by the resonant impedance sheet as a function of the frequency (analytical and simulated).
%%%%%%%%%%%%%%%%%%%%%%%%%%%%
%%%%%%%%%%%%%%%%%%%%%%%%%%%%
The absorbing substance is composed of $\epsr=1.5,\muur=1,\sig=0.05\left(\rm{S}/m\right)$ with $d=0.4\left(m\right)$. In addition, the absorption coefficient of the same absorber (identical electromagnetic parameters and thickness) backed by PEC is presented in dashed red line. It can be observed that although the impedance sheet operates as a PEC at $f_{\rm{res}}$, its large wavelength (small frequencies) capacitive behaviour allows the structure to absorb more efficiently than what is dictated by Rozanov's bound. At the resonance frequency, the absorption coefficient of the resonant structure is identical to the PEC backed absorber, resulting in an intersection point between the blue line and the red-dashed line in Fig.~\ref{Resonantor} (d). The minimal absorption point, i.e $\tilde{A} \sim 0$, is located nearby, but not exactly at the resonance frequency, due to the contribution of the lossy layer. An overlap between these two points occurs when the resonance frequencies of the sheet and the structure (Dallenbach layer terminated by the sheet) are identical, i.e., the resonance frequency is sufficiently small such that the lossy TL can be described effectively as a shunt admittance.

\section{Summary and Conclusion}
%%%%%%%%%%%%%%%%%%
In this manuscript we augmented Rozanov's bound for Dallenbach layers that are backed by partially transparent impedance sheets. We demonstrated analytically and by simulations potential realistic designs of absorbing layers that can absorb beyond what is expected by Rozanov's bound from a layer with the same thickness.
The key point of our approach was obtained by observation on the mathematics of Berenger's PML layers. There, as discussed in Sec.~\ref{magnetic_conductive_beyond}, an additional degree of freedom in the form of magnetic conductivity is used in order to manipulate the long wavelength behaviour of the reflection coefficient, which in turn enables to achieve ultrabroadband absorption.
Taking this observation into realistic designs, we show in Secs.~\ref{Impedance_Boundary_beyond} and \ref{Tightness}, that the introduction of impedance sheets, either inductive, capacitive, or resonant, may yield similar control on the long wavelength behaviour of the reflection coefficient, and consequently on the net absorption. These designs, as shown in the manuscript, provide the possibility for net absorption that is not constrained by Rozanov's bound, albeit in a passive, LTI absorbing layer.
Our findings may be useful in practical designs of absorbers in a wide range of frequencies and physical realms.

\section*{Acknowledgment}
C. F.  would like to thank to the Darom Scholarships and High-tech, Bio-tech and Chemo-tech Scholarships and to Yaakov ben-Yitzhak Hacohen excellence Scholarship. This research was supported by the Israel Science Foundation (grant No. 1353/19).

\appendix
{
\section{Derivation of Eqs.~\eqref{rho_sigma_sigmam}--\eqref{Eq_B}} \label{AppendixA}
%%%%%%%%%%%%%%%%%%%%%
The input impedance as seen at the surface of the lossy medium towards the absorbing layer is given by $Z_{in}(\lambda)=Z_0(\lambda)\tanh(\gamma(\lambda)d)$, with $\gamma(\lambda)=(j/\lambda)\sqrt{(2\pi c_0 \hat{\muu}-j\lambda \hat{\sig}_m)(2\pi c_0 \hat{\eps}-j\lambda\hat{\sig})}$ as the complex propagation term where $\{\hat{\eps},\hat{\muu},\hat{\sig},\hat{\sig}_m\}$ are frequency dependent permittivity, permeability, electric conductivity and magnetic conductivity, respectively. $Z_0(\lambda)$ is the characteristic impedance of a lossy TL which is given by \cite{Collin1966Foundations,Pozar2011Microwave} with its low frequency approxiation($f\rightarrow0$),
%%%%%%
\begin{align}\label{Characteristic_impedance}
  &Z_0(\lambda) =\sqrt{\frac{2\pi c_0 \hat{\muu}-j\lambda\hat{\sig}_m}{2\pi c_0 \hat{\eps}-j \lambda \hat{\sig}}} \sim\sqrt{\frac{\sigmam}{\sig}} \biggl[1+ j\pi c_0   \\
  &\left(\frac{\muu}{\sigmam}-\frac{\eps}{\sig}\right) \lambda^{-1}-\frac{1}{2}\pi^2 c_0^2 \left(3\frac{\eps^2}{\sig^2}-2\frac{\eps }{\sig}\frac{\muu}{\sigmam}-\frac{\muu^2}{\sigmam^2}\right)\lambda^{-2}\biggr]\nonumber
\end{align}
%%%%%%
Similarly, approximating  $\tanh(\gamma(\lambda)d)$ by
%%%%%%%
\begin{align}
\label{Hyperbolic_tangent_approx}
    &\tanh(\gamma(\lambda)d)\sim \tanh(\sqrt{\sigmam\sig}d) \nonumber \\
    & +j\pi c_0 \sqrt{\sigmam\sig}d \left(\frac{\muu}{\sigmam}+\frac{\eps}{\sig}\right) \sech^2(\sqrt{\sigmam\sig}d)\lambda^{-1}  \\
    &+\pi^2 c_0^2 d\sqrt{\sig\sigmam}\sech^2\left(\sqrt{\sig\sigmam}d\right)\biggl[\frac{1}{2} \left(\frac{\muu}{\sigmam}-\frac{\eps}{\sig}\right)^2 \nonumber \\
    &+d \sqrt{\sig \sigmam}\left(\frac{\muu}{\sigmam}+\frac{\eps}{\sig}\right)^2 \tanh(\sqrt{\sig\sigmam}d)\biggr]\lambda^{-2},
    \nonumber
\end{align}
%%%%%%%
{where $\epsilon,\mu,\sigma$ and $\sigma_m$ are the ststic values of the parameters.}
Using, Eqs.~\eqref{Characteristic_impedance} and \eqref{Hyperbolic_tangent_approx}, gives the long wavelength (low frequency) approximation the input impedance,
%%%%%%%
 \begin{align}
     \label{Input_Impedance}
    &Z_{in}(\lambda)\sim \sqrt{\frac{\sigmam}{\sig}}\tanh(\sqrt{\sigmam\sig}d)+j\pi c_0 \sqrt{\frac{\sigmam}{\sig}} \nonumber \\
    &\biggl[ d\sqrt{\sig\sigmam}\left(\frac{\muu}{\sigmam}+\frac{\eps}{\sig}\right) \sech^2(\sqrt{\sig\sigmam}d) \\
    &+ \left(\frac{\muu}{\sigmam}-\frac{\eps}{\sig}\right) \tanh(\sqrt{\sig\sigmam}d) \biggr]\lambda^{-1}+\mathcal{O}(\lambda^{-2}).
    \nonumber
\end{align}
%%%%%%%
The reflection coefficient at the interface between the absorbing layer and the surroundings (the semi-infinite TL) is defined in Eq.~\eqref{reflection}.
Substituting Eq.~\eqref{Input_Impedance} into Eq.~\eqref{reflection} yields the long wavelength approximation of the reflection coefficient in Eq.~\eqref{rho_sigma_sigmam}.
%%%%%%%%
% \begin{equation}
% \label{Reflection_coeff_low}
% \rho(\lambda)\sim -1+A+2\pi jB/\lambda.
% \end{equation}
%%%%%%%%
%where
%%%%%%%%
%\begin{align}
%\label{Reflection_coeff_low_2}
%A&=\frac{2\kappa}{1+\kappa},
%\qquad \qquad
%\kappa=\frac{1}{\eta_0} \sqrt{\frac{\sigma_m}{\sigma}}\tanh\left({\sqrt{\sigma\sigma_m}}d\right),
%\\[1ex]
%B&=c_0\frac{d\left(\frac{\mu}{\sigma_m}+\frac{\epsilon}{\sigma}\right) \left(\frac{\sigma_m}{\eta_0}-\eta_0 \sigma \kappa^2\right)+\left(\frac{\mu}{\sigma_m}-\frac{\epsilon}{\sigma}\right)\kappa}{(1+\kappa)^2}.
%%B&=\frac{c_h}{\eta_0\eta_h}\frac{\frac{d}{\sigma_s}\left(\epsilon_s\sigma_{m_s}+\mu_s\sigma_s\right)\sech^2\left({\sqrt{\sigma_s\sigma_{m_s}}}d\right)+\sqrt{\frac{\sigma_{m_s}}{\sigma_s}}\left(\frac{\mu_s}{\sigma_{m_s}}-\frac{\epsilon_s}{\sigma_s}\right)\tanh\left({\sqrt{\sigma_s\sigma_{m_s}}}d\right)}{\left(1+\frac{\sqrt{\sigma_{m_s}}\tanh\left({\sqrt{\sigma_s\sigma_{m_s}}}d\right)}{\sqrt{\sigma_s}\eta_0\eta_h}\right)^2}
%\label{Reflection_coeff_low_3}
%\end{align}
%%%%%%%%
Note that by applying the small argument approximation $d \ll \delta$, with $\delta=1/\sqrt{\sigmam\sig}$ in Eqs.~\eqref{rho_sigma_sigmam}--\eqref{Eq_B} {(where $\delta$ may be identified as a ``static penetration depth'')}, $\tanh(\sqrt{\sigmam\sig}d) \sim \sqrt{\sigmam\sig}d - (\sigmam\sig)^{3/2}d^3/3$ which with, $\kappa \sim \sigmam d/\eta_0$ and further setting $\sigma_m \to 0$}  identifies with the  Rozanov's approximation \cite{rozanov2000ultimate} $\rho(\lambda)\sim-1+4\pi j \muur d/\lambda$.
%
%While for the ``no electric conductivity'', $\sigma \to 0$, it follows that $\kappa \sim \sigma_m d/\eta_0 \left (1- d^2 \sigma_m \sigma/3\right)$ and  $\rho(\lambda)$ is approximated by,
% \begin{equation}
% \label{Reflection_coeff_noelectric}
% \rho(\lambda)\sim-1 +\frac{\frac{2\sigma_m d}{\eta_0}}{1+\frac{\sigma_m d}{\eta_0}}+\frac{4\pi j \mu_r d}{\lambda\left(1+\frac{\sigma_m d}{\eta_0}\right)^2}\left[1-\frac{\epsilon_r}{3\mu_r}\left(\frac{\sigma_m d}{\eta_0}\right)^2\right].
%\end{equation}
%%%%%%%%%%%%%%%%%%%%%%

{
%%%%%%%%%%%%%%%%%%%%%
\section{Reflection coefficient sum rules}\label{AppendixB}
\label{Complex_Process}
%%%%%%%%%%%%%%%%%%%%%
Here, it is assumed a time dependence of $e^{j2\pi f t}$, with $f$ the frequency and $\lambda=c_0/f$ the wavelength, thus the reflection coefficient $\rho$ is an analytic function in the lower half of the complex $f$-plane or equivalently in the upper half of complex $\lambda$-plane. In view of the long wavelength approximation, $\rho(\lambda)\sim-1 + A+j2\pi B/\lambda$ in Eq.~\eqref{rho_sigma_sigmam}, define $\rho_a(\lambda)=\rho\left(\lambda\right)-A$. $\rho_a(\lambda)$ is also an analytic function in the upper half complex $\lambda$ plane with the long wavelength approximation $\rho_a(\lambda)~\sim -1 + j2\pi B/\lambda$. Recalling Rozanov's bound  derivation in \cite{rozanov2000ultimate} for analytic functions in the upper half of the $\lambda$-plane with the asymptotic behaviour in Eq.~\eqref{Eq_Rozanov_Ref} it follows that this same derivation can be used for $\rho_a$, resulting in Eq.~\eqref{Generalized_Relation}. Here, the result of the integration may be positive, therefore the inequality sign cannot be reversed in contrast to Rozanov's case.
}

\section{Monte Carlo simulations}\label{AppendixE}
\begin{figure}[H]
     \centering
    \begin{subfigure}[b]{0.48\columnwidth}
         \centering
         \includegraphics[width=\textwidth]{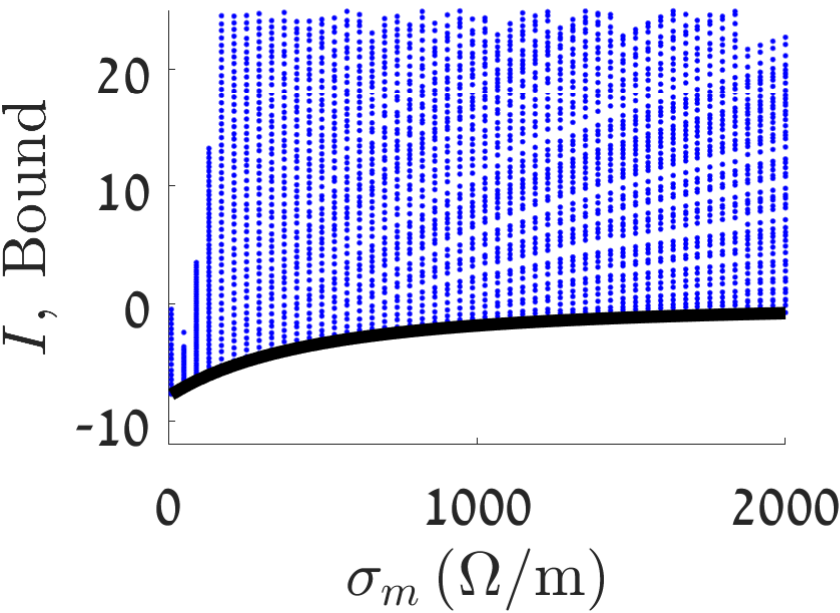}
         \caption{}
         \label{Example_Magnetic}
    \end{subfigure}
    \hfill
    \begin{subfigure}[b]{0.48\columnwidth}
         \centering
         \includegraphics[width=\textwidth]{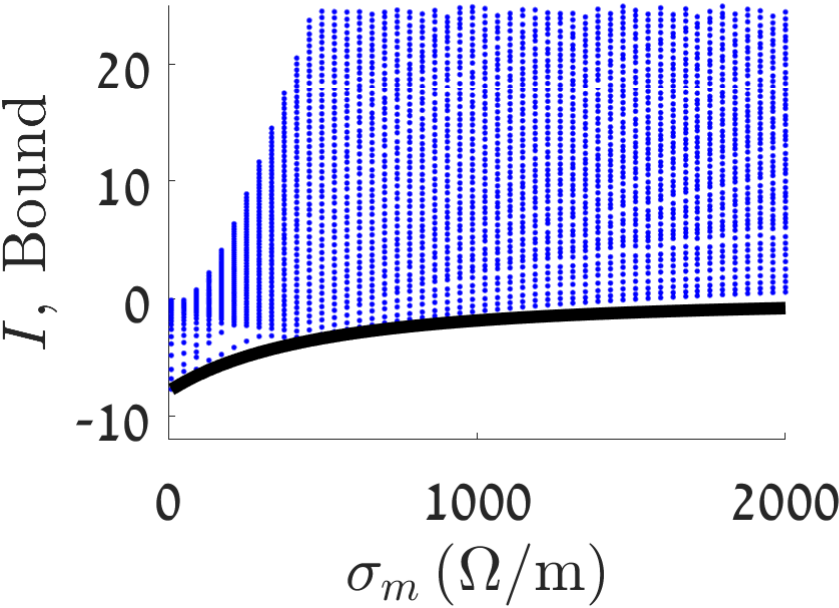}
         \caption{}
         \label{Unit_Cell_Pspice_1}
    \end{subfigure}
    \\
    \begin{subfigure}[b]{0.48\columnwidth}
         \centering
         \includegraphics[width=\textwidth]{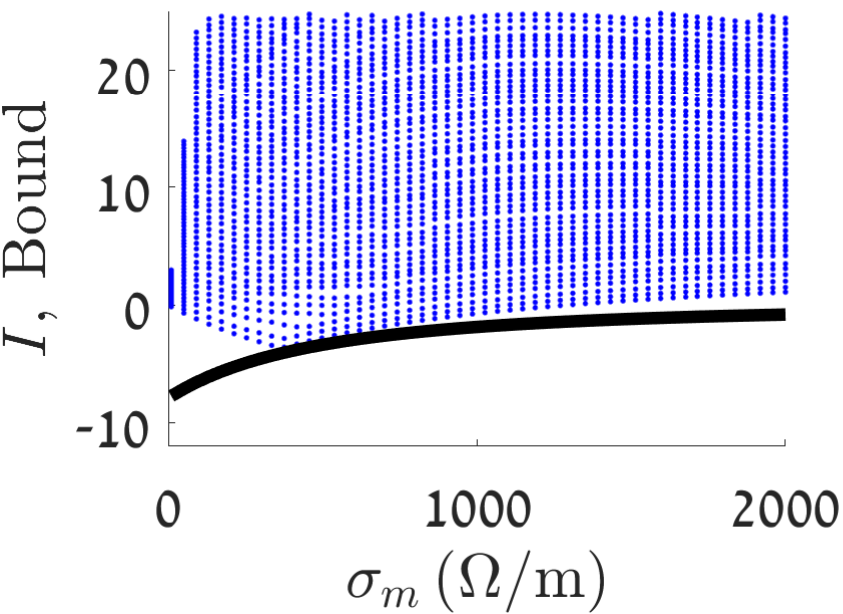}
         \caption{}
         \label{Unit_Cell_Pspice_2}
    \end{subfigure}
    \begin{subfigure}[b]{0.48\columnwidth}
         \centering
         \includegraphics[width=\textwidth]{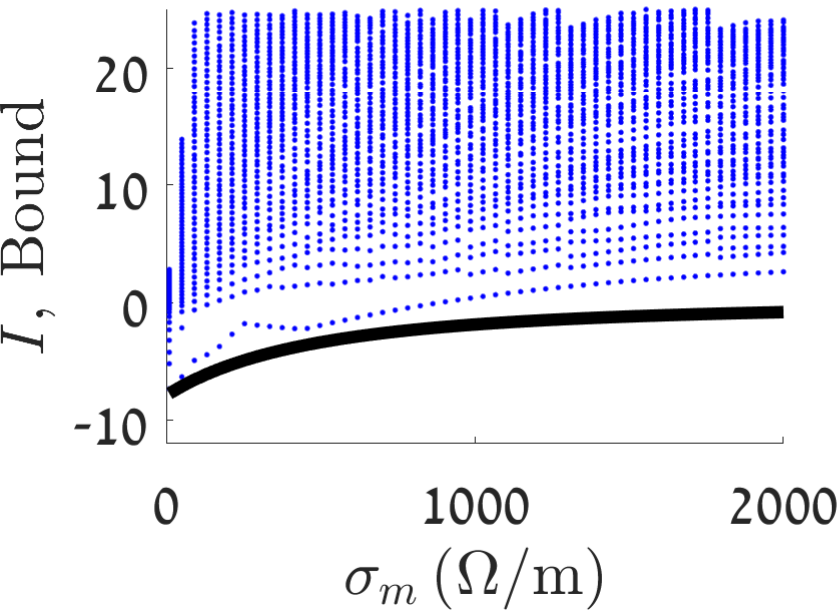}
         \caption{}
         \label{Unit_Cell_Pspice}
    \end{subfigure}
\caption{Monte Carlo simulations of Eq.~\eqref{Generalized_Relation_2} where $\muur=1$. The lower bound, i.e., the RHS of the equation is shown in black, while the integration in the LHS is shown in blue dots for random sets of parameters. (a) The parameters $\{\eps,\muu,\sigmam\}$ were considered as constant over the entire frequency spectrum and equal to the static parameters. For each $\sigmam$, multiple positive values of $\eps$ were randomly generated. (b) For dispersive dielectric constant, we consider a Lorentzian form of the permittivity, $\eps \left(\lambda\right)$, where its parameters, $\{A,\lambda_{\rm{rel}},\lambda_{\rm{res}}\}$ were randomly selected with positive values. {(c),(d) Depict similar results as appear in (a),(b) for wavelength dependent magnetic conductivity. (c) $\epsr=1.5$ (arbitrarily selected). (d) $\epsr \left(\lambda\right)$ has a Lorentzian form given in Eq.\eqref{Relative_permittivity}.}}
\label{Mote-Carlo}
\end{figure}

The bound described in Eq.~\eqref{Generalized_Relation} can be simplified for Dallenbach layers with no static electric conductivity using Taylor series {and in} the limit $\sig\rightarrow0$ (see Appendix \ref{AppendixB}),
\begin{equation}\label{Generalized_Relation_2}
    \int_{0}^{\infty} \ln |\rho(\lambda)-\hat{A}| \,d\lambda \geq\frac{-2\pi^2\muur d}{\left(1+\frac{\sigmam d}{\eta_0}\right)^{2}},\quad \hat{A}=\frac{\frac{2\sigmam d}{\eta_0}}{1+\frac{\sigmam d}{\eta_0}}.
\end{equation}
Here, we perform numerical simulations to verify the bound in Eq.~\eqref{Generalized_Relation_2}, with the absorbing layer having unit relative permeability, $\muur=1$. It can be observed that the right hand side in Eq.~\eqref{Generalized_Relation_2} is independent of $\eps$ and thus only dependent on $\sigmam$. {We considered several scenarios}; in the first, the parameters $(\eps,\sigmam)$ were taken as constant over the entire {wavelength} spectrum and thus equal to the static parameters. For each selection of magnetic conductivity, we generated random positive relative permittivity, $\eps\sim U \left(0,500\right) \times \eps_0$ ($U(x_1,x_2)$ denotes uniform distribution between in the range $(x_1,x_2)$) and evaluated both sides of Eq.~\eqref{Generalized_Relation_2}. Next, we considered a more realistic case where the permittivity has some frequency dependence {($\hat{\eps}$)}. We used the following Lorentzian model \cite{Rakic1998Optical},
\begin{equation}\label{Relative_permittivity}
    \hat{\eps}_r(\lambda)= \eps_{\infty} +\frac{A}{1+j\frac{\lambda_{\rm{rel}}}{\lambda}-\left(\frac{\lambda_{\rm{res}}}{\lambda}\right)^2},
\end{equation}
where $\eps_{\infty}=1$ is the relative permittivity at extremely large frequency and $\lambda_{\rm{res}}$ and $\lambda_{\rm{rel}}$ are the resonance wavelength and damping coefficient, respectively. The positive values of $\{A,\lambda_{\rm{res}},\lambda_{\rm{rel}}\}\sim U\left(0.1,100\right)$ were randomly selected. Let us denote the LHS of Eq.~\eqref{Generalized_Relation_2} {which was calculated numerically for each set of random parametres $\{A,\lambda_{\rm{res}},\lambda_{\rm{rel}}\}$} with $I$ and the RHS will be termed `Bound'. Figure~\ref{Mote-Carlo} depicts in black line the Bound and the blue dots denote $I$ {(for each set of the parameters)}. {Figure~\ref{Mote-Carlo}(a) depicts the constant frequency variation parameters while Fig.~\ref{Mote-Carlo}(b) depicts Lorentzian model}.
It can be observed that the {the value of the numerical integration, $I$,}
in the LHS of Eq.~\eqref{Generalized_Relation_2} is always above the corresponding calculated lower bound, thus indicating that indeed $\eps$ does not affect the bound.
{To enrich the discussion, we considered a wavelength dependent magnetic conductivity in accordance to Drude model for ordinary electric conductors \cite{Jackson1999Classical},
\begin{equation}\label{Drude_Magnetic}
    \hat{\sig}_m \left(\lambda\right)=\frac{\sigmam}{1+j\frac{\lambda_d}{\lambda}},
\end{equation}
where $\lambda_d \sim U\left(0.1,100\right)$ is the magnetic conductivity damping coefficient. Figure~\ref{Mote-Carlo}(c) depicts the simulation results for constant frequency dependence of the permittivity (with $\epsilon_r=1.5$) and Fig.~\ref{Mote-Carlo}(d) presents the results for a Lorentzian dependence as appears in Eq.\eqref{Relative_permittivity}. It can be observed that the RHS of Eq.~\eqref{Generalized_Relation_2} bounds from bellow the LHS.}
%%%%%%%%%%%%%%%%%%%%%%%%%%%%
%%%%%%%%%%%%%%%%%%%%%%%%%%%%

%%%%%%%%%%%%%%%%%%%%%%%%%%%%
\section{Kramers–Kronig relations and wideband constant parameters within a finite frequency band}\label{AppendixD}
Kramers-Kronig relations are mathematical expressions that link the real and imaginary parts of a causal function, which is analytically defined in half of the complex plane. These relations state that in order to evaluate the real (imaginary) part of a complex function at a specific frequency, one must sum over the entire frequency band of the imaginary (real) part \cite{Jackson1999Classical}.
In Section \ref{Tightness}, we focus on modeling realizable systems with practically constant effective parameters within a finite frequency band, such as $f\in(0,2)(\rm{GHz})$. It is important to note that assuming practically constant permittivity or conductivity in this frequency band does not contradict the Kramers-Kronig relations. This is because the resonances of the material occur at much higher frequencies.

\begin{figure}[H]
     \centering
    \begin{subfigure}[b]{0.49\columnwidth}
         \centering
         \includegraphics[width=\textwidth]{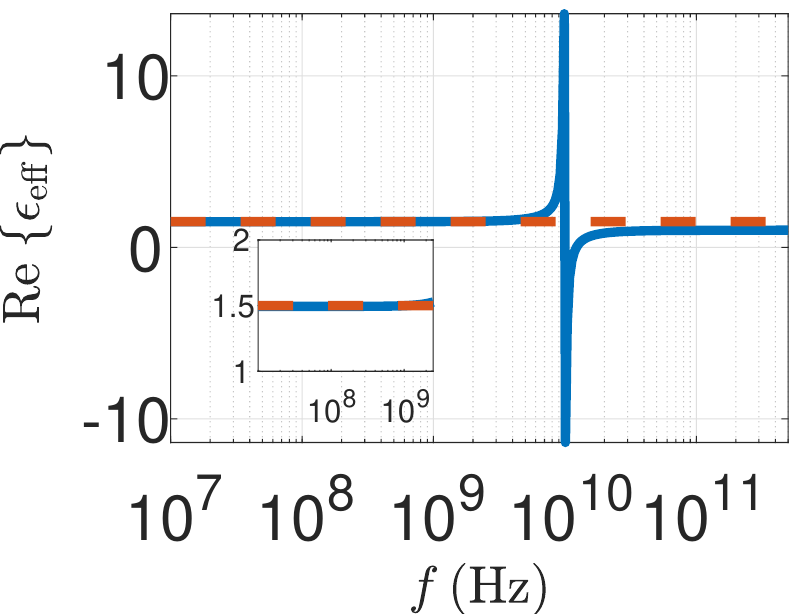}
         \caption{}
         \label{KK_real}
    \end{subfigure}
    \hfill
    \begin{subfigure}[b]{0.49\columnwidth}
         \centering
         \includegraphics[width=\textwidth]{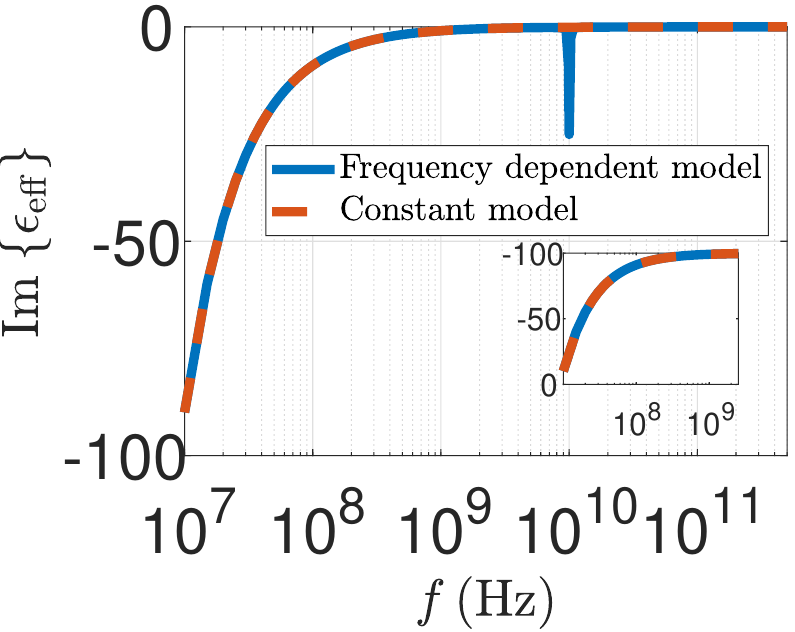}
         \caption{}
         \label{KK_Im}
    \end{subfigure}
\caption{The complex effective permittivity of a lossy substance, taking into account both polarization and conduction loss mechanisms. The blue line represents a causal substance, while the dashed red line depicts a constant model. The parameters in the causal model were carefully chosen to achieve a high level of agreement within the frequency range of $f\in(0,2)(\rm{GHz})$. The real part of the permittivity is shown in (a), while the imaginary part is displayed in (b).}
\label{KK}
\end{figure}

To visually demonstrate this argument, we provide details of a Lorentzian model for the complex permittivity, assuming a time dependence of $e^{j2\pi f t}$. The equation for the complex relative effective permittivity, $\eps_{\rm{eff}}(f)$, is given by \cite{Jackson1999Classical},
\begin{equation}\label{eps_eff}
\begin{split}
&\eps_{\rm{eff}}(f)=\eps_{r}(f)+\frac{\sig(f)}{j2\pi f \eps_0}=\\
&1+\frac{A}{1+\frac{jf}{f_{\rm{rel}}}-(f/f_{\rm{res}})^2}+\frac{\sig}{j2\pi f \eps_0 (1+j2\pi f \tau_{\sig})},
\end{split}
\end{equation}
where $\eps_{\rm{eff}}(f)$ is the complex relative effective permittivity, $A$ is the dielectric strength response, $f_{\rm{rel}}=c_0/\lambda_{\rm{rel}}$  is the relaxation frequency, $f_{\rm{res}}=c_0/\lambda_{\rm{res}}$ is the resonant frequency and $\tau_{\sig}$ is the time constant of the conductor. The first and second terms in the right hand side of Eq.~\eqref{eps_eff} represent the polarization response of the material while the third term describes the conduction mechanism (Drude model).
In our paper, as depicted in Figures \ref{Tighten_Bound} - \ref{Resonantor}, we aim to maintain a constant permittivity and electric conductivity within a finite frequency range. To achieve this, specifically for $f\in(0,2)(\rm{GHz})$, we set the free parameters of the model as follows: $A=0.5$, $f_{\rm{rel}}=5\times 10^{11}, (\rm{Hz})$, $f_{\rm{res}}= 10^{10}, (\rm{Hz})$, $\tau_{\sig} = 10^{-12}, (\rm{sec})$, and $\sig=0.05 (\rm{S/m})$.
We then compare the complex effective relative permittivity, $\eps_{\rm{eff}}(f)$, of the causal model to the desired "constant" model with $\eps_r=1.5$ and $\sig=0.05, (\rm{S/m})$. The results are presented in Figure \ref{KK}, where the blue line represents the causal model and the red dashed line corresponds to the desired "constant" model. An very good agreement is observed within the desired frequency range. However, it is worth noting that further improvement in agreement can be achieved by positioning the resonance at an even higher frequency band.

%%%%%%%%%%%%%%%%%%%%%%%%%%%%
%%%%%%%%%%%%%%%%%%%%%%%%%%%%

%%%%%%%%%%%%%%%%%%%%%%%%%%%%

%\bibitem{}

\end{document}